\documentclass{article}

\usepackage{arxiv}
\usepackage{float}
\usepackage{relsize}
\usepackage{ulem}
\usepackage[utf8]{inputenc} 
\usepackage[T1]{fontenc}    
\usepackage{hyperref}       
\usepackage{url}            
\usepackage{booktabs}       
\usepackage{amsfonts}       
\usepackage{nicefrac}       
\usepackage{mathrsfs}
\usepackage{microtype}      
\usepackage{lipsum}
\usepackage{subcaption} 
\usepackage{algpseudocode}
\usepackage{algorithm, algorithmicx}
\usepackage{graphicx}
\usepackage{color}
\usepackage[dvipsnames]{xcolor}
\usepackage{amsmath}
\usepackage{amssymb}
\usepackage{amsthm}
\usepackage{multirow}
\usepackage[backend=biber,doi=true,url=true,isbn=true,giveninits=true]{biblatex}
\AtBeginBibliography{\footnotesize} 
\graphicspath{ {./images/} }

\theoremstyle{plain}
\newtheorem{theorem}{Theorem}[section]
\newtheorem{lemma}{Lemma}[section]

\theoremstyle{definition}

\theoremstyle{remark}
\newtheorem{remark}{Remark}[section]

\title{Fast automatically differentiable matrix functions and applications in molecular simulations}

\author{
 Tina Torabi\thanks{These authors contributed equally to this work.} \\
  Department of Mathematics\\ 
  University of British Columbia\\ 
  Vancouver, V6T1Z2, BC, Canada \\
  \texttt{torabit@student.ubc.ca} \\
   \And
Timon S. Gutleb\footnotemark[1] \\
  School of Computer Science\\ 
  University of Leeds\\ 
  Leeds, LS2 9JT, UK \\
\texttt{T.S.Gutleb@leeds.ac.uk} \\
  \And
  Christoph Ortner \\
  Department of Mathematics\\ 
  University of British Columbia\\ 
  Vancouver, V6T1Z2, BC, Canada \\
  \texttt{ortner@math.ubc.ca} \\
}

\makeatletter
\def\@fnsymbol#1{\ifcase#1\or \dagger\fi}
\makeatother

\begin{document}

\maketitle

\begin{abstract}
We describe efficient differentiation methods for computing Jacobians and gradients of a large class of matrix functions including the matrix logarithm $\log(A)$ and $p$-th roots $A^{\frac{1}{p}}$. 
We exploit contour integrals and conformal maps as described by (Hale et al., SIAM J. Numer. Anal. 2008) for evaluation and differentiation and analyze the computational complexity as well as numerical accuracy compared to high accuracy finite difference methods. 
As a demonstrator application we compute properties of structural defects in silicon crystals at positive temperatures, requiring efficient and accurate gradients of matrix trace-logarithms. 
\end{abstract}

\section{Introduction}
Matrix functions have long been integral to mathematical and scientific disciplines and thus have a wide range of applications \cite{alma99139172010001452, doi:10.1137/1.9780898717778, alma99161779544601452} in fields such as network analysis \cite{https://doi.org/10.1002/gamm.202000012}, control systems \cite{article}, matrix function neural networks \cite{batatia2024equivariantmatrixfunctionneural}, and solid state physics \cite{doi:10.1137/15M1022628, julian2016, torabi2024surrogate}.
For instance, in materials science, matrix functions are indispensable for calculating vibrational entropy, which provides insights into the impact of thermal vibrations on material properties \cite{torabi2024surrogate, julian2016}.

Despite their broad applicability, the computation of derivatives of matrix functions—in particular for matrix families \(X_\theta\) parameterized by a vector \( \theta \)—remains a significant computational challenge. This challenge is particularly acute in high-dimensional systems, where conventional methods often become computationally prohibitive. For instance, in the study of thermally activated processes \cite{CrystalPlasticity} precise calculations of free energy derivatives are required. Conventional approaches often rely on potential energy surface (PES) approximations, which, while computationally feasible, can introduce significant errors when entropy plays a major role. To accurately capture the full behavior of such systems, it is necessary to consider the free energy surface (FES) instead, which requires the computation of more complex matrix function derivatives.

In this paper we introduce a method that leverages analytic functional calculus in conjunction with both forward and reverse mode automatic differentiation (AD). Our objective is to develop a robust framework that significantly enhances the efficiency and scalability of computing derivatives of matrix functions. Our approach streamlines computational implementation and expands the potential applications of matrix function derivatives to more complex higher-dimensional problems.

\section{Matrix functions} \label{sec:Matrixfs}
\subsection{Analytic matrix functions}
We use the term \textit{matrix function} to refer to a matrix-valued extension of a scalar function \( f \) such that \( f(X) \) with \( X \in \mathbb{C}^{n \times n} \) is a computable matrix retaining the dimensions of \( X \). If $f(x) = a_0 + a_1 x + \dots + a_K x^K$ is a polynomial, this extension is given by 
\[
    f(X) = \sum_{k = 0}^K a_k X^k,
\]
with $X^0 = I_n$, the $n \times n$ identity matrix. For general continuous  $f$, $f(X)$ is defined via density of polynomials. Several approaches for evaluating matrix functions have been considered in the literature~\cite{doi:10.1137/1.9780898717778}, e.g. via the eigendecomposition (or more generally the Jordan canonical form), or via a specific polynomial interpolation scheme. In this work we are primarily concerned with evaluating analytic matrix functions.  

Suppose that $f: \mathbb{C} \rightarrow \mathbb{C}$ is analytic on a closed set $\overline{\mathcal{D}} \subset \mathbb{C}$, where $\mathcal{D}$ is an open set containing the spectrum of a matrix $X \in \mathbb{C}^{n \times n}$; i.e., $\sigma(X) \subset \mathcal{D}$. Assume further that the boundary, $\mathcal{C} := \partial\mathcal{D}$, is comprised of a finite number of closed rectifiable Jordan curves, encircling $\sigma(X)$ once in the counterclockwise direction. Then $f(X)$ can be expressed as a Riemann contour integral over $\mathcal{C}$ as follows: 
\begin{equation} 
    f(X)= \frac{1}{2 \pi i} \oint_{\mathcal{C}} f(z) \; (zI - X)^{-1} \,{\rm d} z. \label{eq:cauchy}
\end{equation}

\subsection{Fast computation of matrix functions \(f(X)\)} \label{sec:Halemethod}
In physics and engineering applications, performant numerical evaluation of matrix functions is often essential. Hale et al. \cite{doi:10.1137/070700607} demonstrated that the Cauchy integral definition in \eqref{eq:cauchy} is particularly useful for numerical computations. They described an efficient approach utilizing \eqref{eq:cauchy} for matrices \( A \) whose eigenvalues are located on or near the positive real axis \((0, \infty)\), and for functions \( f(z) \) such as \( z^\alpha \) or \( \log z \), which are analytic except for the presence of singularities or a branch cut near the negative real axis \((- \infty, 0]\). This section reviews their method which we use in subsequent sections to also compute derivatives of matrix functions.

\begin{figure}[!htb]
    \centering
    \begin{subfigure}[b]{0.48\textwidth}
        \centering
\includegraphics[width=\textwidth]{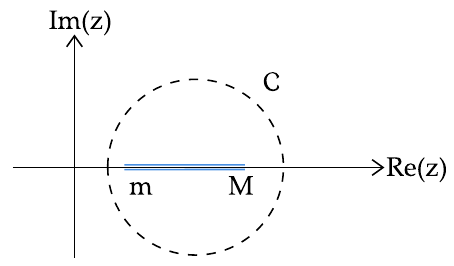}
        \caption{Spectrum of \(X\)}
        \label{fig:contour-init}
    \end{subfigure}
    \hspace{0.5in}
    \begin{subfigure}[b]{0.3\textwidth}
        \centering
\includegraphics[width=\textwidth]{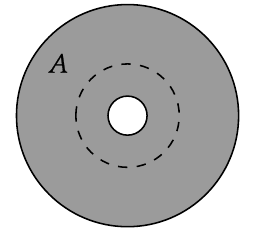}
        \caption{Annulus of analyticity}
        \label{fig:annulus}
    \end{subfigure}
    \caption{Figure (a) depicts the spectrum of $X$ along with the appropriate contour $\mathcal{C}$. Figure (b) demonstrates mapping the entire doubly connected domain of analyticity $\Xi$, onto an annulus region $X$ through a conformal map. In this figure, the interval of singularities corresponds to the outer boundary of the annulus, while the interval including the spectrum aligns with the dashed inner boundary circle. We then apply the trapezoid rule over a circle in the annulus.}
\end{figure}

Consider a matrix \(X\) with real entries and eigenvalues in \((0, \infty)\) as illustrated in Figure \ref{fig:contour-init}. While \(X\) is symmetric in many important cases, facilitating the application of trapezoidal sums, symmetry is not essential for the methods presented and will thus not be assumed. We will refer to the minimum and maximum eigenvalues of \(X\) as \( \rm m\) and \(\rm M\), respectively. In cases where \(X\) is either symmetric or normal, the 2-norm condition number is given by the ratio \( \rm M/m\). We will generally assume that \( \rm m\) and \(\rm M\) are known though we note that they would typically only be estimated in practice. We further assume that the spectrum \(\sigma(X)\) completely fills the interval \([\rm m, \rm M]\), without significant gaps and that the function \(f\) is analytic in the slit complex plane \(\mathbb{C} \setminus (-\infty, 0]\).
At first glance, one might think of surrounding \([ \rm m,  M] \) by an appropriate contour, such as depicted in Figure \ref{fig:contour-init}, followed by applying the trapezoid rule to approximate the integral. However, for ill-conditioned matrices this is inefficient as it necessitates \( O(M/m) \) linear solves to achieve typical accuracy requirements. The techniques suggested in \cite{doi:10.1137/070700607} leverage variable transformations and conformal mappings to optimize the choice of contour points, effectively reducing the computational complexity to \( O(\log(M/m)) \).

\begin{figure}
     \centering
        \includegraphics[width=0.9\textwidth]{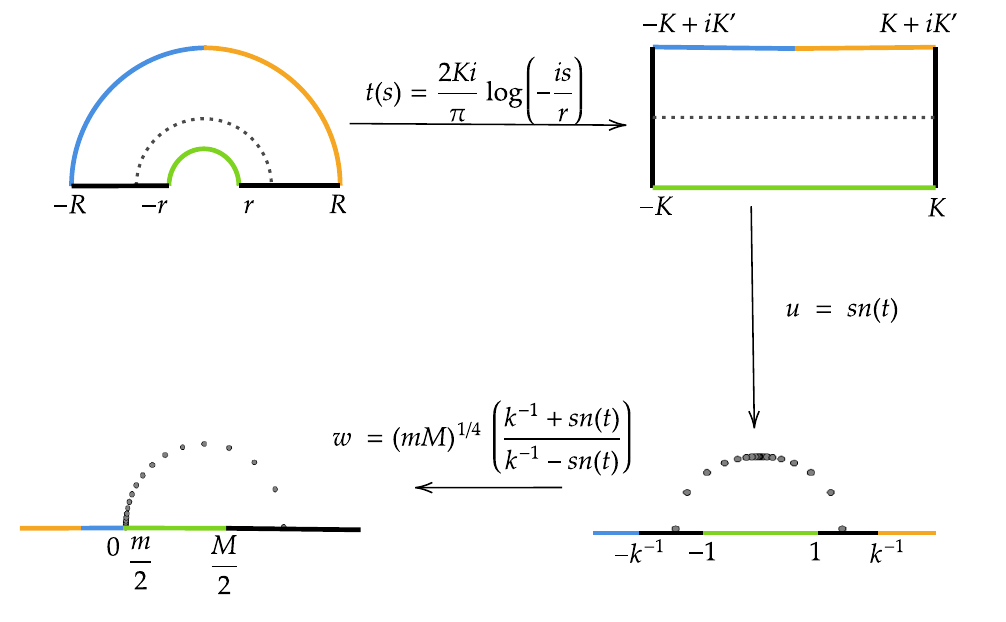}
    \caption{Steps of the conformal map shown in detail.} 
    \label{fig:fullsteps}
\end{figure}

The conformal mapping employed involves multiple transformations designed to map the region of analyticity of \( f \) and \( (z I - X)^{-1} \), characterized by the doubly connected set \( \Xi = \mathbb{C} \setminus ((-\infty, 0] \cup [\frac{m}{2}, \frac{M}{2}]) \), onto an annulus \( A = \{ z \in \mathbb{C} : r < |z| < R \} \), where \( r \) and \( R \) represent the inner and outer radii of the annulus respectively. We show a schematic of this transformation in Figure \ref{fig:annulus}. The question of how to map $\Xi$ to the annulus $A$ and vice versa was answered by Hale et al. \cite{doi:10.1137/070700607} in three steps: Starting in the $s$-plane, we first map the annulus to a rectangle with vertices $\pm K$ and $\pm K+iK'$, using a logarithmic transformation
\begin{equation}
t(s) = \frac{2Ki}{\pi} \log\left(-\frac{is}{r}\right),
\end{equation}
where $K, K'$ denote the complete elliptic integrals. For more details on Jacobi elliptic functions and integrals, see Appendix \ref{app:JEF}. Next, the rectangle is mapped to the upper half-plane in the $u$-plane by the Jacobian elliptic function
\begin{equation}
u(t) = \text{sn}(t|k^2), \quad k = \frac{(M/m)^{1/2} - 1}{(M/m)^{1/2} + 1}.
\end{equation}
A Möbius transformation is then applied to map the upper half-plane to the $z$-plane
\begin{equation}
z(u) = (M/m)^{1/2} \left(\frac{k^{-1} +u}{k^{-1} - u} \right).
\end{equation}
This final transformation is designed to distribute the eigenvalues of $X$ evenly along the real axis, thus facilitating the application of the trapezoidal rule~\cite{doi:10.1137/070700607}. The steps of this conformal map are shown in Figure \ref{fig:fullsteps}.
The integral in \eqref{eq:cauchy} can be reformulated as follows:
\begin{equation}
f(X) = -\frac{X}{2\pi i} \int_{-K+iK'/2}^{3K+iK'/2} f(z(t)) (z(t) - X)^{-1} z^{-1} \, \frac{{\rm d}z}{{\rm d}u} \, \frac{{\rm d} u}{{\rm d}t} \, {\rm d}t,
\end{equation}
where the interval from \(-K + iK'/2\) to \(K + iK'/2\) reflects the segment of \(\Gamma\) in the upper half-plane, extended to \(3K + iK'/2\) to include the lower half-plane contribution, with \(z\) transformations provided by:
\begin{equation}
\frac{dz}{du} = \frac{2k^{-1}\sqrt{mM}}{(k^1 - u^2)^2}, \quad \frac{du}{dt} = \text{sn}(t) = \sqrt{1 - k^2u^2} = \text{cn}(t) \text{dn}(t).
\end{equation}
Here, \(\text{cn}\) and \(\text{dn}\) are standard Jacobi elliptic functions \cite[{(22.2.4-22.2.6)}]{NIST:DLMF}. 
Therefore, the expression for \(f(X)\) becomes:
\begin{equation}
f(X) = -\frac{X\sqrt{mM}}{\pi k} \int_{-K+iK'/2}^{3K+iK'/2} \frac{f(z(t)) (z(t) - X)^{-1} z^{-1}\text{cn}(t) \text{dn}(t)}{(k^{-1} - u)^2} \, {\rm d}t.
\end{equation}
Given that \(X\) is real,
the integrand is real-symmetric, implying that \(f(X)\) is effectively double the real part of the integral evaluated over the first half of the contour. Simplifying, we have:
\begin{equation}
f(X) = -\frac{2X\sqrt{mM}}{\pi k} \; \text{Im} \int_{-K+iK'/2}^{K+iK'/2} \frac{f(z(t)) (z(t) - X)^{-1} z^{-1}\text{cn}(t) \text{dn}(t)}{(k^{-1} - u)^2} \, {\rm d}t.
\end{equation}

Applying the trapezoidal rule, with \(t_j = -K + \frac{iK'}{2} + 2 \frac{(j-1/2)K}{N}\) for \(j = 1, \ldots, N\), representing \(N\) equidistant points in \(-K + iK'/2, K + iK'/2\), results in

\begin{equation}
f_N(X) = -\frac{4KX \sqrt{mM}}{\pi Nk} \text{Im} \left( \sum_{j=1}^N f(z(t_j)) \frac{(z(t_j) I - X)^{-1} \text{cn}(t_j) \text{dn}(t_j)}{z(t_j)(k^{-1} - u(t_j))^2} \right).
\end{equation}

The convergence rate of the numerical approximation of \( f(X) \) using an \( N \)-point quadrature formula to the true value of \( f(X) \) is given in~\cite[Theorem 1]{doi:10.1137/070700607}. Specifically, the error between the actual function \( f(X) \) and the numerical approximation \( f_N(X) \) is bounded by:

\[
\|f(X) - f_N(X)\| = O\left(e^{-\pi^2 N / \log(M/m + 3)}\right).
\]

It is noteworthy that when $f$ has a singularity at \(z=0\) but just a branch cut on \((-\infty,0)\), the above mentioned method is not as efficient as in the case when $f$ has singularities on \((-\infty,0)\). 
To reach fast convergence, Hale et al. \cite{doi:10.1137/070700607} proposed a change of variable $w = \sqrt{z}$, ${\rm d}z=2w\,{\rm d}w$, under which \eqref{eq:cauchy} becomes

\begin{equation}
    f(X) = \frac{X}{\pi i} \int_{\mathcal{C}_w} w^{-1} f(w^2) (w^2 - X)^{-1} {\rm d}w.
\end{equation}

In the revised approach, the branch cut of the function \( f(z) \) along the negative real axis in the \( z \)-plane is unfolded to the imaginary axis in the \( w \)-plane, allowing \( f(z) \) to be analytically continued as \( f(w^2) \) across the entire slit \( w \)-plane, i.e., \(\mathbb{C} \setminus (-\infty, 0]\). The method involves a contour integration within a modified region, specifically enclosing \([m^{1/2}, M^{1/2}]\) in the \( w \)-plane. This adaptation improves the domain from \([m, M]\) to \([m^{1/2}, M^{1/2}]\), enabling a more effective application of the previous computational techniques.

The expression used for this refined method is
\begin{equation}
f_N(X) = -\frac{8K X \sqrt[4]{mM}}{\pi Nk} \text{Im} \left( \sum_{j=1}^N f(w(t_j)^2) \frac{(w(t_j)^2 I - X)^{-1} \text{cn}(t_j) \text{dn}(t_j)}{w(t_j) (k^{-1} - u(t_j))^2} \right), \label{eq:cauchy_num}
\end{equation}
where the \( t_j \) are specified by the modified equations, reflecting changes in the contour parameters:
\[
w = \sqrt[4]{mM} \frac{k^{-1} + \text{sn}(t)}{k^{-1} - \text{sn}(t)}, \quad k = \frac{\sqrt[4]{M/m} - 1}{\sqrt[4]{M/m} + 1}.
\]
These adjustments allow for precise integration over the new contour, efficiently capturing the integral's value while minimizing computational overhead.
\subsection{Derivatives of matrix functions}
The framework for evaluating matrix functions \( f(X) \) reviewed in Section~\ref{sec:Halemethod} can be efficiently extended to families of matrices denoted as \( X(u) \), where \( u \in \Omega \subset \mathbb{R}^m \) represents a multi-dimensional parameter, as well as derivatives of \( f(X(u)) \) with respect to the parameter \( u  \in \Omega \). Here, $\Omega \subset \mathbb{R}^m$ is an open parameter domain and $m \in \mathbb{N}$ the number of parameters. 

\begin{lemma}
    Consider a matrix \( X(u) \), where each entry \( x_{ij}: \Omega \to \mathbb{C} \) is continuously differentiable. Assume that \( f \) is analytic in a domain \( \bar{D} \), where \( D \) is open and includes the spectrum of \( X(u) \) for all \( u \in \Omega \). Then:

    The derivatives of the matrix function \( f(X) \) with respect to the component \( u_k \) of \( u \) are given by:
    \[
    \frac{\partial f(X)}{\partial u_k} = \frac{1}{2\pi i} \int_{\mathcal{C}} f(z) (zI - X)^{-1} \left(\frac{\partial X}{\partial u_k}\right) (zI - X)^{-1} \, \mathrm{d}z,
    \]
    where \( \mathcal{C} \) is a positively oriented contour enclosing the spectrum of \( X \), and \( zI \) represents the identity matrix scaled by \( z \).
    \label{lemma:der}
\end{lemma}

The proof based on \eqref{eq:cauchy} is elementary and hence omitted.

Computing the derivatives \(\frac{\partial f(X)}{\partial u_m}\) with respect to each component \(u_m \) of a multi-dimensional parameter vector \( u \) can lead to substantial computational complexity since it involves evaluating \(\frac{\partial X}{\partial u_m}\),  the Jacobian of \( X \) with respect to \( u \). The resulting Jacobian \(\frac{\partial X}{\partial u}\) is a tensor of dimensions \( n \times n \times m \), representing a significant computational bottleneck. In the section which follows, we discuss the computational complexity of this approach in more detail and explain how reverse mode differentiation avoids this in many important scenarios.

\section{Computational complexity} \label{sec:compcomplexity}

\subsection{Objective and Formulation} 

\begin{figure}[!htb]
    \centering
    \begin{subfigure}[b]{0.48\textwidth}
        \centering
\includegraphics[width=\textwidth]{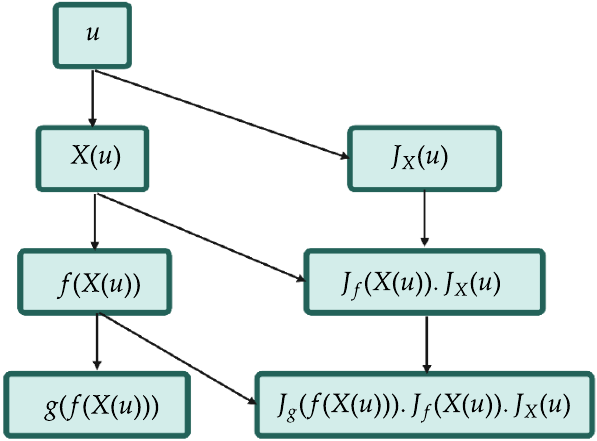}
        \caption{Computational graph for the forward mode.}
        \label{fig:fwd}
    \end{subfigure}
    \hspace{0.1in}
    \begin{subfigure}[b]{0.48\textwidth}
        \centering
\includegraphics[width=\textwidth]{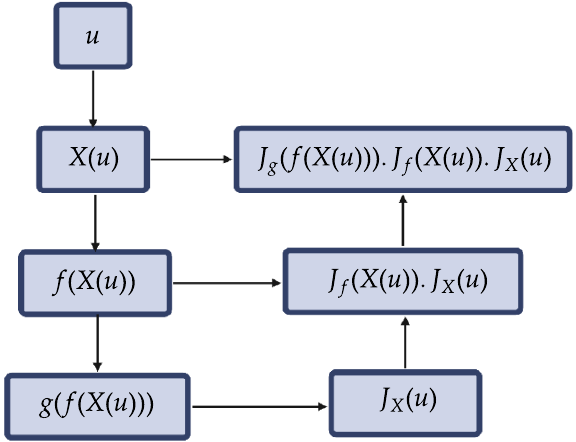}
        \caption{Computational graph for the reverse mode.}
        \label{fig:rev}
    \end{subfigure}
    \caption{Computational graphs for the forward and reverse modes of differentiation for the composite function \( h(u) = g(f(X(u))) \). }
    \label{fig:compgraph}
\end{figure}

Let \( g : \mathbb{R}^{n \times n} \to \mathbb{R}^p \), \( f: \mathbb{R}^{n \times n} \to \mathbb{R}^{n \times n} \), and \( X:\mathbb{R}^{m} \to \mathbb{R}^{n \times n} \) be differentiable functions. Define the composite function \( h : \mathbb{R}^m \to \mathbb{R}^p \) by \( h(u) = g(f(X(u))) \), where the parameter vector \( u \in \mathbb{R}^m \) is mapped to a \( p \)-dimensional output. As illustrated in Figure~\ref{fig:compgraph}, our goal is to evaluate the sensitivity of the output \( h \) with respect to changes in the input \( u \), which is captured by the Jacobian \( J_h(u) = \displaystyle\tfrac{\partial h}{\partial u} \). This Jacobian quantifies how infinitesimal changes in the input vector \( u \) affect the output \( h(u) \), and can be computed by systematically applying the chain rule to differentiate through the layers of the composite function. Applying the chain rule and elementary matrix multiplication, we can express the Jacobian as:
\begin{equation}
J_h(u) = J_{g \circ f \circ X}(u) = J_g(f(X(u))) \cdot J_f(X(u)) \cdot J_X(u)
\label{eq:jacrep}
\end{equation}


This chain rule can be applied in two distinct ways depending on how we propagate the derivatives: either by starting from the inputs and working towards the outputs (\textit{forward mode differentiation}) shown in Figure~\ref{fig:fwd}, or by starting from the outputs and working back towards the inputs (\textit{reverse mode differentiation}) shown in Figure~\ref{fig:rev}. Although the chain rule is the same in both cases, the direction in which it is applied leads to different computational complexities.  To analyze the computational complexity of evaluating the Jacobian \( J_h(u) \) in reverse mode, we must first understand the costs associated with the forward evaluations of each function in the composition \( h(u) = g(f(X(u))) \). Specifically, we denote the cost of evaluating each function without differentiation as \(\texttt{cost}(X)\), \(\texttt{cost}(f)\), and \(\texttt{cost}(g)\), respectively.


Our aim is to compute these derivatives via computer programs. In order to evaluate numerical derivatives rather than symbolic expressions, automatic differentiation (AD)~\cite{hoffmann2016, doi:10.1137/1.9780898717761, Griewank2012OnTN, HechtNielsen1989TheoryOT, VANDENBERG2024103010} computes derivatives by accumulating intermediate values during code execution. Standard code may be easily modified to incorporate AD with little effort. 
The target function \(h \) must first be constructed as a series of basic operations in a computer program before applying AD. After that, AD can be used in either forward or reverse mode.

\subsection{Review: Forward mode differentiation via dual numbers}
To begin, we will briefly go over dual numbers and their applications in forward mode AD. Dual numbers~\cite{Gu1987DualnumberTA, ADpiponi} extend the real numbers by introducing an infinitesimal unit \( \epsilon \) with the property
\[
\epsilon^2 = 0, \quad \epsilon \neq 0.
\]
A dual number can be written in the form:
\[
x = a + b\epsilon,
\]
where \( a, b \in \mathbb{R} \), \( a \) is called the real part, and \( b \) is called the dual part. The Taylor series expansion of a function \( f \) evaluated at a dual number \( a + b\epsilon \) around \( a \) is given by:
\[
f(a + b\epsilon) = f(a) + f'(a)b\epsilon + \frac{f''(a)}{2}(b\epsilon)^2 + \frac{f'''(a)}{6}(b\epsilon)^3 + \cdots.
\]

Since the property \( \epsilon^n = 0 \) holds for $n \geq 2$, the Taylor series simplifies to:
\[
f(a + b\epsilon) = f(a) + f'(a)b\epsilon.
\]


This simplification highlights the utility of dual numbers in computing first-order derivatives. Thus, dual numbers offer a straightforward and reliable approach for computing derivatives. There exist highly performant practical implementations of dual numbers, which simply extend basic arithmetic operations and a short list of standard mathematical functions from standard floating point numbers to dual numbers.
For example, the \texttt{ForwardDiff.jl}~\cite{RevelsLubinPapamarkou2016}  package in Julia and the cppduals~\cite{Tesch2019} library in C++ provide efficient and well-documented implementations of dual number arithmetic.

Suppose we want to compute the directional derivative of a function \( h \) with respect to \( m' \) parameters of the input vector \( x \in \mathbb{R}^m \). To achieve this, \textit{forward mode} AD~\cite{10.5555/3122009.3242010, Wengert1964ASA, Griewank1989OAD} utilizes \( m' \) dual numbers, where each dual number, \( \epsilon_i \), represents an independent infinitesimal perturbation associated with the direction \( v_i \in \mathbb{R}^m\). Specifically, it evaluates

\begin{equation}
    h\bigg( x + \overset{m'}{\underset{i=1}{\sum}} \epsilon_i v_i\bigg)
    = h(x) + \sum_{i = 1}^{m'} \epsilon_i \nabla h(x)\cdot v_i.
\end{equation}

In the dense case, this approach is conceptually equivalent to computing $m'$ directional derivatives independently. However, in sparse cases efficient techniques can be used to reduce the computational cost. More generally, forward mode AD computes the action of the Jacobian \( J_h(x) \) on a matrix \( V \) of size \( m \times m' \), where the columns of \( V \) are the \( m' \) directional vectors:
\[
V \mapsto J_h(x)V.
\]
This is often referred to as the pushforward, adapted from differential geometry terminology. Forward mode AD computes these directional derivatives by following the process shown in Figure~\ref{fig:fwd}, where derivatives are propagated step by step:


\begin{table}[H]
    \centering
    \begin{tabular}{|c|l|}
        \hline
        \textbf{Computation} & \textbf{Cost} \\ \hline
        \(\displaystyle (v_1, v_1' ) = \big(X(c), J_X(c) V\big)\) 
        & \(m' \cdot \texttt{cost}(X)\) \\ \hline
        \(\displaystyle (v_2, v_2') = \big(f(v_1), J_f(v_1) \cdot v_1'\big)\) 
        & \(m' \cdot \texttt{cost}(f)\) \\ \hline
        \(\displaystyle (v_3, v_3') = \big(g(v_2), J_g(v_2) \cdot v_2'\big)\) 
        & \(m' \cdot \texttt{cost}(g)\) \\ \hline
    \end{tabular}
    \caption{Computational steps and their associated costs during each step of forward mode AD.}
    \label{tab:fwd_sweep_costs}
\end{table}

Computing the full Jacobian, corresponds to computing all $m$ directional derivatives and essentially starting with $V$ as the identity matrix $I_m$. The total cost of evaluating \( J_h(c) \) is then $m' \cdot (\texttt{cost}(X) + \texttt{cost}(f) + \texttt{cost}(g))$ and thus depends on the methods chosen to compute the functions $X, f$ and $g$. We will discuss this further in the next subsection.

\subsection{Review: Reverse mode differentiation}
In contrast to forward mode, \textit{reverse mode} differentiation~\cite{doi:10.1137/1.9780898717761, doi:10.1137/1.9781611971200} computes derivatives by propagating information backward from the output(s) of a function. This mode computes \textit{adjoints} (or sensitivities) with respect to the outputs, rather than computing directional derivatives with respect to the inputs and is particularly advantageous for functions where the output dimension is smaller than the input dimension. Given \( Y \in \mathbb{R}^{p \times p} \), in the cotangent space, reverse mode AD computes the action of the transposed Jacobian on \( Y \):
\begin{equation}
    J_h(u)^\top Y, 
\end{equation}
and computes the sensitivities with respect to $p$ outputs, which essentially construct the rows of $J_h(u)$. Once the forward evaluation is complete as shown in Figure~\ref{fig:rev}, the adjoints are propagated in reverse as follows:

\begin{table}[H]
    \centering
    \begin{tabular}{|c|l|}
        \hline
        \textbf{Computation} & \textbf{Cost} \\ \hline
        \(\nabla_f = J_g(f(X(u)))^\top \cdot Y\) & \(p \cdot \texttt{cost}(g)\) \\ \hline
        \(\nabla_X = J_f(X(u))^\top \cdot \nabla_f = \delta_X[f^\top \cdot \nabla_f]\) 
        & \(p \cdot \texttt{cost}(f)\) \\ \hline
        \(\nabla_u = J_X(u)^\top \cdot \nabla_X = \delta_u[X^\top \cdot \nabla_X]\) 
        & \(p \cdot \texttt{cost}(X)\) \\ \hline
    \end{tabular}
    \caption{Computational steps and their associated costs during each sweep of reverse mode AD.}
    \label{tab:rev_sweep_costs}
\end{table}

In reverse mode, computing the full Jacobian involves evaluating the sensitivities with respect to all \( p \) outputs. This is achieved by initializing \( Y \) as the identity matrix $I_{p}$. Consequently, the total cost of computing \( J_h(c) \) in reverse mode is given by: $p\cdot (\texttt{cost}(X) + \texttt{cost}(f) + \texttt{cost}(g))$, where the overall cost naturally again depends on the methods used to compute the functions $X$, $f$ and $g$. The choice of the most efficient mode of AD (forward or reverse) depends on comparing the number of outputs \( p \) and the number of inputs \( m \). Reverse mode is more efficient when \( p \ll m \), while forward mode is better suited for cases where \( p \gg m \).

Additionally, evaluating the Jacobian of matrix functions \( f(X(u)) \) with respect to the parameter vector \( u \), denoted as \( J_f(u) = J_f(X(u)) J_X(u) \), is of significant importance. In forward mode, this computation aligns with the first two rows of Table~\ref{tab:fwd_sweep_costs}, and the total cost of evaluating the full Jacobian scales as \( m \cdot ( \texttt{cost}(f) + \texttt{cost}(X)) \). 

In reverse mode, the computation corresponds to the last two rows of Table~\ref{tab:rev_sweep_costs}, with the total cost scaling as \( n^2 \cdot (\texttt{cost}(f) + \texttt{cost}(X)) \). Clearly, unless $n^2<m$,
reverse mode Jacobian evaluation is significantly more computationally expensive than forward mode. The corresponding complexities for evaluating $J_f(u)$ are summarized in Table~\ref{tab:comp_jac}. 



It is noteworthy that advanced techniques such as sparsity detection and matrix coloring can be employed to further reduce the computational cost of evaluating the Jacobian, by efficiently identifying and exploiting the structure within the matrices involved in the differentiation process. For a detailed description of this we refer to Appendix \ref{app:coloring}. A numerical demonstration of this approach will be provided in section~\ref{sec:jacobian}.



\subsection{Simplified Cost Analysis}To estimate the cost of evaluating \( f(X) \), we revisit the methods described in Section~\ref{sec:Matrixfs}. The computational cost depends on the size of the matrix \( X \), its structural properties (e.g., sparsity or bandedness), and the number of quadrature points \( \ell \) used in the evaluation. If we take \( g = \mathrm{Trace} \), we can rewrite \( (zI - X) = LU \), and we would then need to evaluate \( \mathrm{Trace}(X (zI - X)^{-1}) = \mathrm{Trace}(X U^{-1}L^{-1}) \). 
Using the Cauchy integral definition~\eqref{eq:cauchy_num}, the cost for different types of matrices is detailed below.

\paragraph{Dense Matrices.} 
For a dense \( n \times n \) matrix \( X \), the cost of evaluating \( f(X) \) scales as \( O(\ell n^3) \), since computing the inverse of a dense matrix and performing other required operations both scale cubically with \( n \). However, in many applications, matrices often exhibit additional structures, such as sparsity or bandedness, which significantly reduce the computational cost. This reduction is achieved because sparse linear solvers can exploit these structural properties to minimize the number of operations required.

\paragraph{Sparse Matrices.}

For a sparse matrix \( X \), the cost of evaluating \( f(X) \) arises from computing
\[
f(X) = \sum_i e_i^\top U^{-1} L^{-1} e_i = \sum_i \left( U^{-\top} e_i \right)^\top  \left( L^{-1} e_i \right),
\] 
where \( U \) and \( L \) are the LU factors of \( X \), and \( e_i \) represents the \( i \)-th standard basis vector. Each term in the summation involves two back-substitutions: one to compute \( L^{-1} e_i \) and another for \( U^{-\top} e_i \). The cost of each back-substitution depends on the number of nonzeros (\(\mathrm{nnz}\)) in the LU factors.

If \( X \) is a banded matrix with a bandwidth \( b \), the LU factorization incurs minimal fill in~\cite{Davis_Rajamanickam_Sid-Lakhdar_2016}. The cost of LU factorization is \( O(nb^2) \). Assuming \( b \) is constant, since the number of nonzeros in $L$ and $U$ factors scale as $O(nb)$, the total cost of evaluating \( f(X) \) scales as \( O(\ell b n^2) \).

For 2D sparse matrices, which often arise from discretizations of 2D grids or the Hessians of (quasi-)two-dimensional structures, LU factorization introduces moderate fill-in. The cost of factorization scales as \( O(n^{3/2}) \), while the number of nonzeros in the \( L \) and \( U \) factors is \( O(n \log n) \). Consequently, the total cost of solving linear systems and evaluating \( f(X) \) scales as \( O(\ell n^2 \log n) \).

For 3D sparse matrices, LU factorization incurs more significant fill-in. The factorization cost scales as \( O(n^2) \), and the number of nonzeros in the \( L \) and \( U \) factors is \( O(n^{4/3}) \). As a result, the total cost of evaluating \( f(X) \) scales as \( O(\ell n^{7/3}) \). 

On the other hand, if eigendecomposition is used to evaluate \( f(X) \), the computational cost is \( O(n^3) \), as the method does not leverage the sparsity of \( X \). Thus, the contour integration approach with its ability to exploit sparsity has clear advantages for large matrices, though for three-dimensional systems that advantage is less pronounced than in lower dimension.

We summarize the computational complexities for each case in Table~\ref{tab:comp}.

\begin{table}[H]
    \centering
    \caption{Complexity comparison for evaluating $J_h(u)$ in the dense and various sparse cases.}
    
    \begin{tabular}{@{}lcccc@{}}
        \toprule
        \textbf{Mode} & \textbf{Sparse 1D} & \textbf{Sparse 2D} & \textbf{Sparse 3D} &
        \textbf{Dense} \\ 
        \midrule
        Forward via contour                    & $O(\ell b m n^2)$        & $O(\ell m n^2 \log n)$ & $O(\ell m n^{7/3})$ & $O(\ell m n^3)$  \\
        Reverse via contour                    & $O(\ell b p n^2)$        & $O(\ell p n^2 \log n)$ & $O(\ell p n^{7/3})$ & $O(\ell p n^3)$ \\
        \addlinespace
        Forward via eigendecomposition           & $O(m n^3)$           & $O(m n^3)$          & $O(m n^3)$ 
        & $O(m n^3)$   \\
        Reverse via eigendecomposition     & $O(p n^3)$         & $O(p n^3)$           & $O(p n^3)$          & $O(p n^3)$ \\
        \bottomrule
    \end{tabular}
    \label{tab:comp}
\end{table}

\paragraph{Improved complexity via selected inversion.} 
The methods we use in our numerical tests have the complexity described above. In principle one can improve the computational cost further through a selected inversion algorithm: To compute \(\mathrm{Trace}(X(zI - X)^{-1})\) for evaluating \( f \) using~\eqref{eq:cauchy_num}, efficient algorithms such as \texttt{Pselinv}~\cite{10.1145/2786977},
which are specifically designed to compute selected elements of a matrix inverse without requiring the full inverse. By leveraging the sparsity and structure of the matrix, \texttt{Pselinv} enables significant computational savings compared to direct inversion. The trace can be expressed as:
\begin{equation}
    \mathrm{Trace}(X(zI - X)^{-1}) = \sum_{i,j} X_{ij} \left((zI - X)^{-1}\right)_{ji},
\end{equation}
where only the entries of \((zI - X)^{-1}\) that correspond to the sparsity pattern of \(X\) are required. Using \texttt{Pselinv}, these entries are computed efficiently, avoiding the computation of the full inverse. The computational cost of \texttt{Pselinv} is comparable to the cost of LU factorization, as it directly builds upon the LU decomposition of the matrix \(zI - X\). For sparse matrices, the cost scales with the number of nonzero elements and the fill-in introduced during factorization. For banded matrices with constant bandwidth, the cost is \(O(n)\), for 2D sparse matrices, the cost is \(O(n^{3/2})\), and for 3D sparse matrices, the cost is \(O(n^2)\). \texttt{Pselinv} is both memory-efficient and highly parallelizable, making it ideal for large-scale problems where sparsity in the matrix can be exploited.

\subsection{Accuracy}

We now turn to the question of accuracy: In general, derivatives computed using automatic differentiation have effectively arbitrarily good precision since the chain rule and explicitly known derivative rules are used. This does not directly apply to our case, however, since the function whose derivative we compute is merely approximated using a contour integral with a finite number of quadrature points. In circumstances where a function evaluation is subject to approximation errors, it is often natural to use approximate derivative schemes such as finite differences schemes of the desired accuracy. It is thus a natural question to ask whether the approximation accuracy of the contour integral methods is sufficient to where use of automatic differentiation is sensible. We will provide numerical experiments to showcase the accuracy in Sections~\ref{sec:num} and ~\ref{sec:app}.

\begin{table}[H]
    \centering
    \caption{Comparison of complexity for evaluating $J_f(u)$ in the dense, sparse 1D, sparse 2D, and sparse 3D cases. When $p \ll n^2$, the cost of $J_f$ is significantly higher than the cost of $J_h(u)$ and evaluating $J_f$ is therefore almost never advantageous.}
    
    \begin{tabular}{@{}lcccc@{}}
        \toprule
        \textbf{Mode}  & \textbf{Sparse 1D} & \textbf{Sparse 2D} & \textbf{Sparse 3D} & \textbf{Dense}\\ 
        \midrule
        Forward via contour                    & $O(\ell b m n^2)$        & $O(\ell m n^2 \log n)$ & $O(\ell m n^{7/3})$ & $O(\ell m n^3)$\\
        Reverse via contour                    & $O(\ell n^4)$        & $O(\ell n^4 \log n)$ & $O(\ell n^{13/3})$ & $O(\ell n^5)$ \\
        \addlinespace
        Forward via eigendecomposition             & $O(m n^3)$           & $O(m n^3)$          & $O(m n^3)$ & $O(m n^3)$  \\
        Reverse via eigendecomposition            & $O(n^5)$           & $O(n^5)$          & $O(n^5)$  & $O(n^5)$  \\
        \bottomrule
    \end{tabular}
    \label{tab:comp_jac}
\end{table}


\section{Numerical Experiments}\label{sec:num}
We present two toy problems to illustrate the complexity and accuracy of the discussed method. Following these examples, we discuss our methodology applied to problems arising from physics, showcasing the method's performance in relevant scenarios.
\subsection{Implementation Details}

We used the {\tt ComplexElliptic.jl} \cite{Torabi2024} package to compute the conformal maps required to implement the contour integral matrix function as described in \cite{doi:10.1137/070700607}. The forward mode AD examples in the toy problems were then computed using  \texttt{ForwardDiff.jl} \cite{RevelsLubinPapamarkou2016} while the reverse mode AD examples used \texttt{Zygote.jl} \cite{Zygote.jl-2018} and \texttt{ChainRules.jl} \cite{FramesWhite2024}. Companion code which reproduces the presented results is available \cite{gitEntropyGrad}.

\subsection{Scalar derivative}

We consider the following toy problem: the composition of two functions \( g(f(X)) \), where the functions \( g \) and \( f \) are defined as 
\[
g(A) = \sum_{i,j} A_{ij}^3, \qquad A = f(X) = X^{1/3}.
\]

The matrix \( X \) is constructed as follows to have connectivity corresponding to a one-dimensional problem setting and to ensure that it is symmetric and positive semi-definite (SPD):  Given a vector \( u \in \mathbb{R}^n \) (i.e., $m = n$), we define 
\begin{align*}
    X(u) &= B(u)^\top B(u), \qquad \text{where} \\ 
    B(u) &= 
    \begin{cases}
        \sin(u_i + u_j) + \cos(u_i u_j) + \exp(u_i - u_j), & \text{if } |i - j| \leq 3, \\
        0, & \text{otherwise}.
        \end{cases}        
\end{align*}

\begin{figure}[!htb]
    \centering
    \begin{subfigure}[b]{0.45\textwidth}
        \centering
\includegraphics[width=\textwidth]{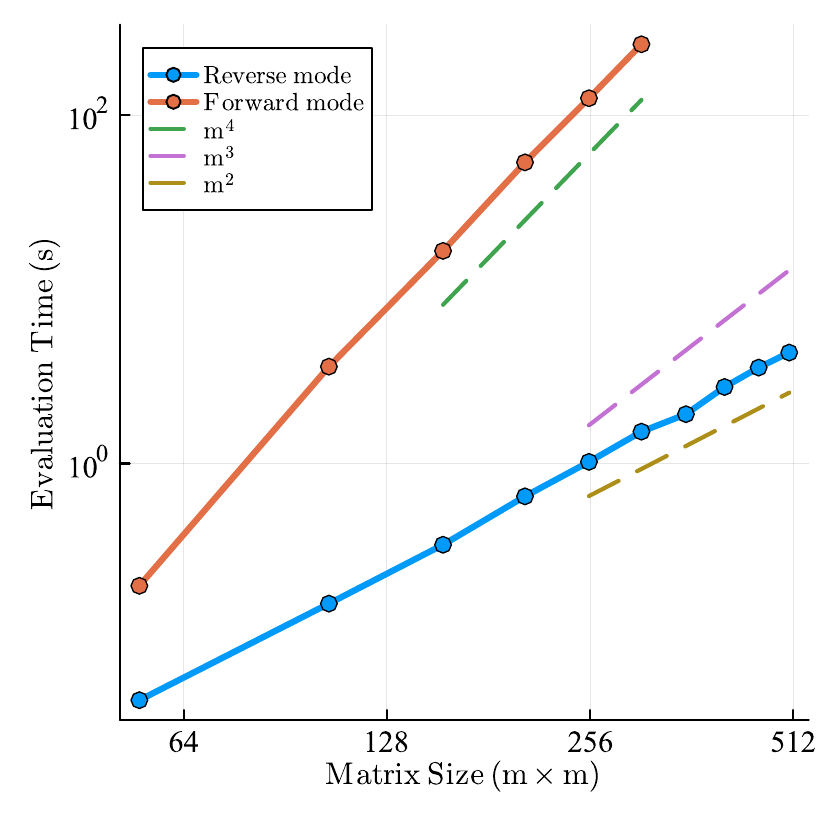}
        \caption{}
        \label{fig:complexityplot_a3}
    \end{subfigure}
    \hspace{0.1in}
    \begin{subfigure}[b]{0.45\textwidth}
        \centering
\includegraphics[width=\textwidth]{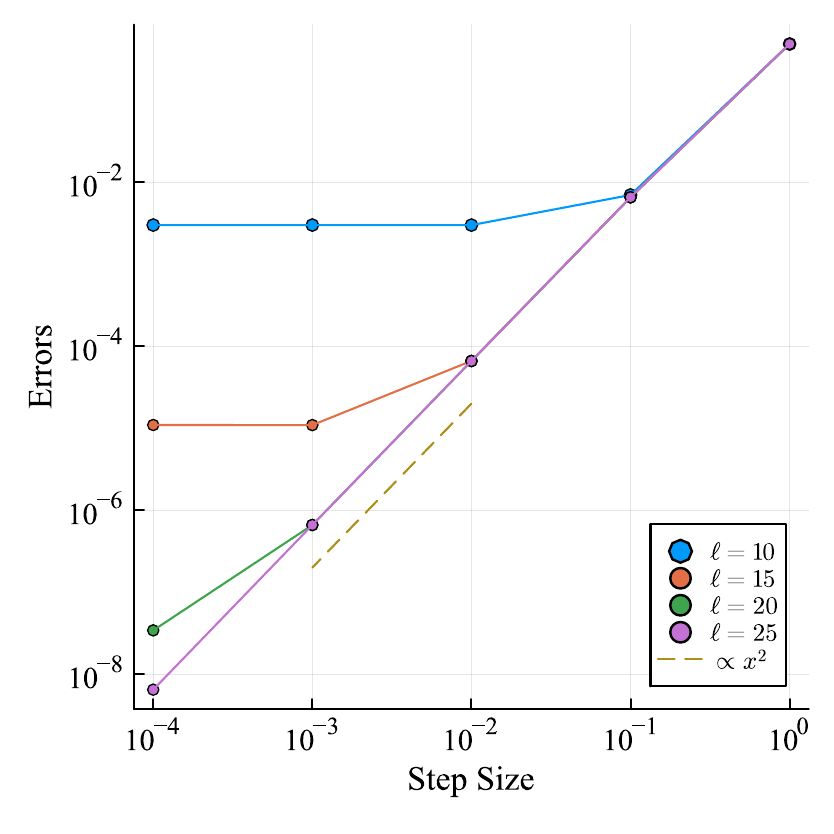}
        \caption{}
        \label{fig:fd_error_a3}
    \end{subfigure}
    \caption{Figure \textbf{(a)} shows obtained evaluation time for computing the gradient of function $g$ with respect to m-vector $u$ using a contour integral quadrature with $\ell=35$ quadrature points. Figure \textbf{(b)} shows convergence of a second order centered finite difference computation of the gradient compared to a gradient computed by reverse AD through the contour integral with conformal map using $\ell$ quadrature points.}
\end{figure}

Figure~\ref{fig:complexityplot_a3} illustrates the evaluation time required to compute the gradient of $g$ with respect to $u$ using a contour integral quadrature with
$\ell=35$ quadrature points. The evaluation times are shown for both reverse mode (blue curve) and forward mode (orange curve) automatic differentiation. 

\begin{remark}
As expected the reverse mode is significantly more efficient as the matrix size increases for scalar derivatives. The observed deviation from the previously mentioned complexity orders is due to the sparsity structure present in the intended applications, which were thus also included in the toy examples. Matrices involved in the differentiation process, here $X$ and $B$, usually exhibit additional structure such as sparsity or even bandedness that reduce the number of operations required \cite{10.1093/acprof:oso/9780198508380.001.0001}.  In particular if as in this example $m=n$, then bandedness with bandwidth $s$ (or equivalent sparsity structure) of the $n^2 \times m$ matrix $J_X(u)$ in Equation~\eqref{eq:jacrep} reduces the expected asymptotic complexity in reverse mode to $O(\ell s n^2)$ in line with our observations.
\end{remark}

Figure~\ref{fig:fd_error_a3} shows the error between the second-order centered finite difference method and reverse AD using the contour integral and conformal map approach, plotted against step size for various quadrature points \(\ell = 10, 15, 20, 25\). The errors decrease as \(\ell\) increases, with higher \(\ell\) values (e.g., \(\ell \geq 25\)) yielding lower errors across all step sizes. The dashed line representing \( \propto x^2 \) indicates the expected quadratic convergence for the finite difference method. The figure also highlights the importance of choosing an appropriate \(\ell\) to balance accuracy and computational cost.

\subsection{Jacobian} \label{sec:jacobian}
Consider the function 
\[
    g(u) := f(X(u)) = \sqrt{ \nabla^2 E(u) }, 
\]
where $X(u) = \nabla^2 E(u)$, $f(X) = \sqrt{X}$ and $E(u), u \in \mathbb{R}^n$,  is given by 
\begin{equation}\label{eq:toymodel}
    E(C; u) = \sum_{i} \left( \sum_{j \in \mathcal{N}_i} C_{ij} |u_i - u_j|^2 + \frac{1}{2}|u_i - u_j|^3 + |u_i - u_j|^4 \right).
\end{equation}
Here, \( i \) indexes discrete points, \( \mathcal{N}_i \) represents the set of nearest neighbors of the \( i \)-th element, \( C_{ij} \) are coefficients characterizing the interactions between \( i \) and \( j \). We can think of $E(u)$ as a toy model for an atomistic potential energy, making this example related to the more realistic cases we consider in the next section. 


\begin{figure}[!htb]
    \centering
    \begin{subfigure}[b]{0.45\textwidth}
        \centering
\includegraphics[width=\textwidth]{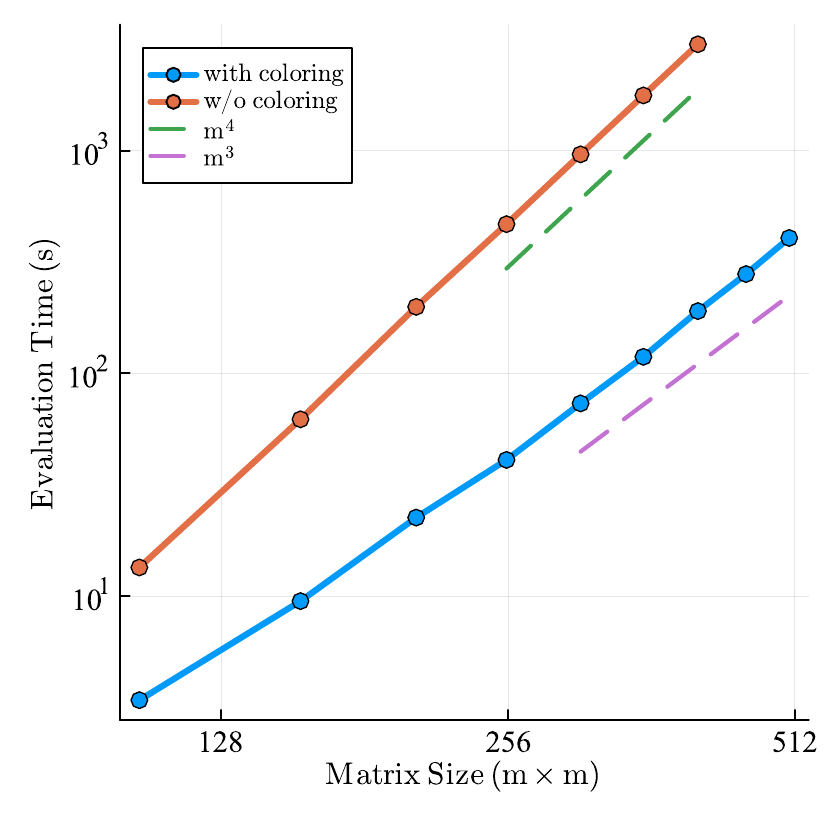}
        \caption{}
        \label{fig:sqrt-complexityplot}
    \end{subfigure}
    \hspace{0.1in}
    \begin{subfigure}[b]{0.45\textwidth}
        \centering
\includegraphics[width=\textwidth]{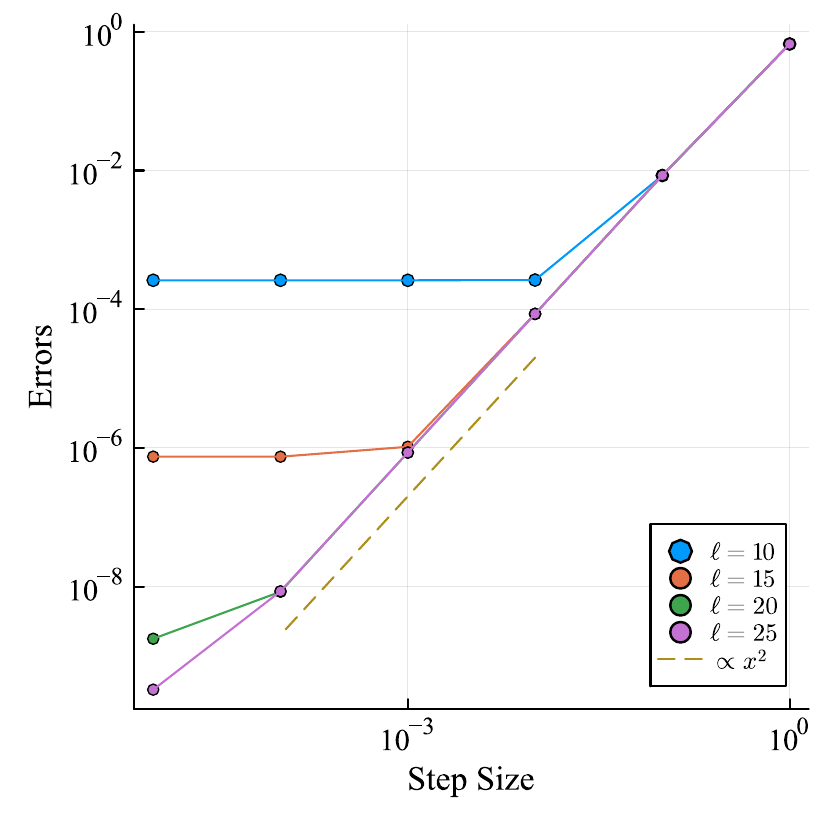}
        \caption{}
        \label{fig:sqrt-errorconv}
    \end{subfigure}
    \caption{Figure \textbf{(a)} shows obtained evaluation time for computing the Jacobian of matrix X with respect to m-vector $u$ using a contour integral quadrature with $\ell=25$ quadrature points. Figure \textbf{(b)} shows convergence of a second order centered finite difference computation of the Jacobian of $f$ compared to one computed by forward mode AD through the contour integral with conformal map using $\ell$ quadrature points.}
\end{figure}

Here, we present the computational complexity of Jacobian evaluation using forward differentiation, comparing the approaches with and without coloring. The details of these methods are provided in Appendix~\ref{app:coloring}. As shown in Figure~\ref{fig:sqrt-complexityplot}, leveraging the sparsity of \( X \) through coloring reduces the Jacobian evaluation cost to \( O(m^3) \), whereas without coloring, the complexity remains at \( O(m^4) \), as expected.

Figure~\ref{fig:sqrt-errorconv} shows the error between the second-order finite difference method and forward mode AD using the contour integral approach across step sizes for various quadrature points \(\ell\), demonstrating that appropriate choice of quadrature points yields highly accurate results.

\section{Application to Defects in Crystalline Silicon}\label{sec:app}

To demonstrate practical applications of our approach, we explore thermally activated processes in materials science, specifically focusing on defect migration, where matrix function derivatives play a crucial role. Understanding these processes is essential for predicting and optimizing the mechanical and electronic properties of materials~\cite{WARNER20094267, Turnbull1951, PhysRevLett.112.225701}.

We consider the free energy 
\begin{align}\label{eq:freeE}
\mathcal{F}(u) = \mathcal{E}(u) - T \mathcal{S}(u),
\end{align}
where \( \mathcal{E}(u) \) denotes the potential energy, \( \mathcal{S}(u) \) represents the entropy, and \( T \) is the temperature. The vibrational entropy \cite{torabi2024surrogate, FULTZ2010247, PhysRevMaterials.6.113803} in solids, \(\mathcal{S}(u)\), is defined as:
\begin{eqnarray}\label{eq:entropy}
\mathcal{S}(u) := -\frac{1}{2} \mathrm{Trace} \log^+(\textbf{F} H(u) \textbf{F}),
\end{eqnarray}
where \( u \) is the lattice displacement vector, and \( H(u) = \nabla^2 \mathcal{E}(u) \) is the Hessian of the potential energy $\mathcal{E}(u)$. To create a spectral gap, we define a self-adjoint operator \( \textbf{F} \), which serves as a preconditioner and acts as \( H_{\text{hom}}^{-1/2} \), where \( H_{\text{hom}} \) is the Hessian of the homogeneous lattice~\cite{torabi2024surrogate, julian2016}. The \(\log^+\) function is defined as follows: let \(\mathbf{T}\) be a bounded, self-adjoint operator on a Hilbert space with spectrum \(\sigma(\mathbf{T}) \subset (-\infty, 0] \cup [m, M]\), where \(0 < m \leq M\). Then, we can define a contour \(\mathcal{C}\) that encircles the interval \([m, M]\) while remaining in the right half-plane. We can then define: 
\begin{equation}
    \log^+(\textbf{T}) = \frac{1}{2\pi i} \oint_{\mathcal{C}} \log (z)\cdot {\rm Trace} \big(  zI- \textbf{T}\big)^{-1}\,{\rm d}z.
\end{equation}

Defect migration is typically analyzed using the potential energy surface (PES), represented by \( \mathcal{E}(\bar{u}) \) in \eqref{eq:freeE}, which governs the structure, dynamics, and thermodynamics of the system \cite{compchem2011}. Migration behavior is often studied by examining stationary points on the PES, such as local minima and transition states, which map out migration pathways, often visualized as steepest-descent paths.

The Nudged Elastic Band (NEB) method \cite{10.1063/1.1323224} is widely used to determine minimum energy paths (MEPs) between known initial and final states on the PES. NEB methods discretize the reaction pathway into a series of intermediate configurations, or `images', connected by elastic springs to ensure even distribution along the path. These images are optimized to remain constrained to the PES while the spring forces maintain equidistant spacing.

While MEPs on the PES are frequently used due to their simplicity, they are only approximations \cite{Zimmerman_2000} and can become increasingly inaccurate at elevated temperatures where entropic effects play an important role. For instance, it has been demonstrated in \cite{PhysRevX.4.011018, PhysRevB.79.134106, Gourav2019, doi:10.1073/pnas.1418241112} that the MEPs on the free energy surface (FES), \( \mathcal{F}(u) \), provides a more accurate and comprehensive depiction of defect migration under finite temperature conditions. By capturing more of the thermodynamic behavior of the system, the FES allows for the analysis of migration mechanisms, rate constants, and material properties that the PES alone cannot adequately describe. Strong entropic effects have also been observed in a variety of processes, such as the nucleation of dislocation loops~\cite{Ryu2010EntropicEO, bagchi2024anomalousentropydrivenkineticsdislocation}, the transformation of vacancy clusters into stacking-fault tetrahedra~\cite{PhysRevLett.99.135501}, the growth of nano-voids under tensile stress~\cite{PhysRevLett.110.206001}, and dislocation emission from crack tips~\cite{WARNER20094267}. In these cases, the role of entropy, particularly at high temperatures, becomes dominant in determining the system's kinetics and overall behavior. Such entropic contributions, while increasingly recognized as fundamental, are impossible to capture using the PES alone, as it only accounts for potential energy minima and transition states, neglecting the broader thermodynamic landscape.

Applying optimization methods such as NEB methods to the free energy surface requires gradients of the free energy functional \( \mathcal{F}(u) \) as expressed in Equation~\eqref{eq:freeE}. These gradients which drive the optimization process necessitate computing derivatives that account for both energetic and entropic contributions. Specifically, the gradient of \( \mathcal{F}(u) \) with respect to the displacement field \( u\) is
\begin{align}\label{eq:freeE_deriv}
\frac{\partial \mathcal{F}(u)}{\partial u} = \frac{\partial \mathcal{E}(u)}{\partial u} - T \frac{\partial \mathcal{S}(u)}{\partial u},
\end{align}
where \( \frac{\partial \mathcal{E}(u)}{\partial u} \) is the derivative of the potential energy, and \( \frac{\partial \mathcal{S}(u)}{\partial u} \) is the derivative of the entropy. The latter involves matrix derivatives that capture changes in entropy due to structural perturbations, which can be computationally expensive.

To address these challenges, various methods have been proposed for exploring the free energy surface in collective variable spaces. Techniques such as metadynamics \cite{Laio2002-bg, CARTER1989472}, adaptive biasing force methods \cite{Dickson2010-eq}, and umbrella sampling \cite{TORRIE1977187} effectively sample the FES in low-dimensional spaces. However, these methods are typically less effective at locating saddle points in high-dimensional spaces, which are critical for accurately determining transition states and reaction pathways in complex systems. These limitations highlight the need for more efficient approaches to calculate free energy derivatives in high-dimensional contexts \cite{10.1063/1.5120372}.

This motivates the development of computationally efficient and highly accurate methods for matrix differentiation to advance our understanding of defect migration and similar thermally activated processes. Next, we thus illustrate the complexity and accuracy of our method for evaluating entropy derivatives and then showcase its application in some practical settings.

\begin{figure}[!htb]
    \centering
    \begin{subfigure}[b]{0.45\textwidth}
        \centering
\includegraphics[width=\textwidth]{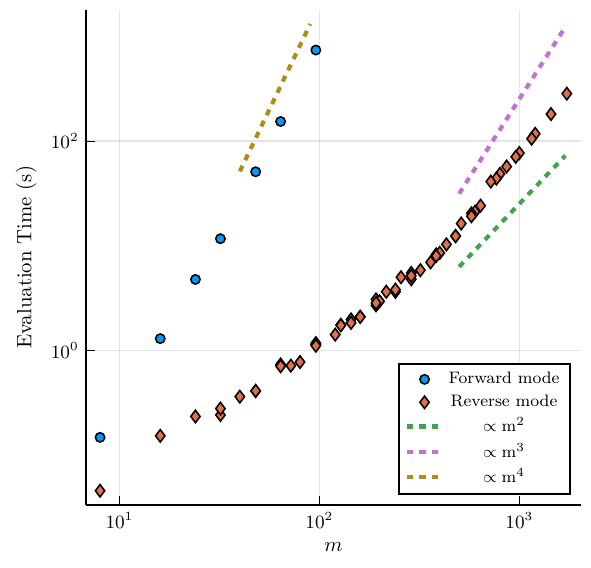}
        \caption{}
        \label{fig:complexityplot_S}
    \end{subfigure}
    \hspace{0.1in}
    \begin{subfigure}[b]{0.45\textwidth}
        \centering
\includegraphics[width=\textwidth]{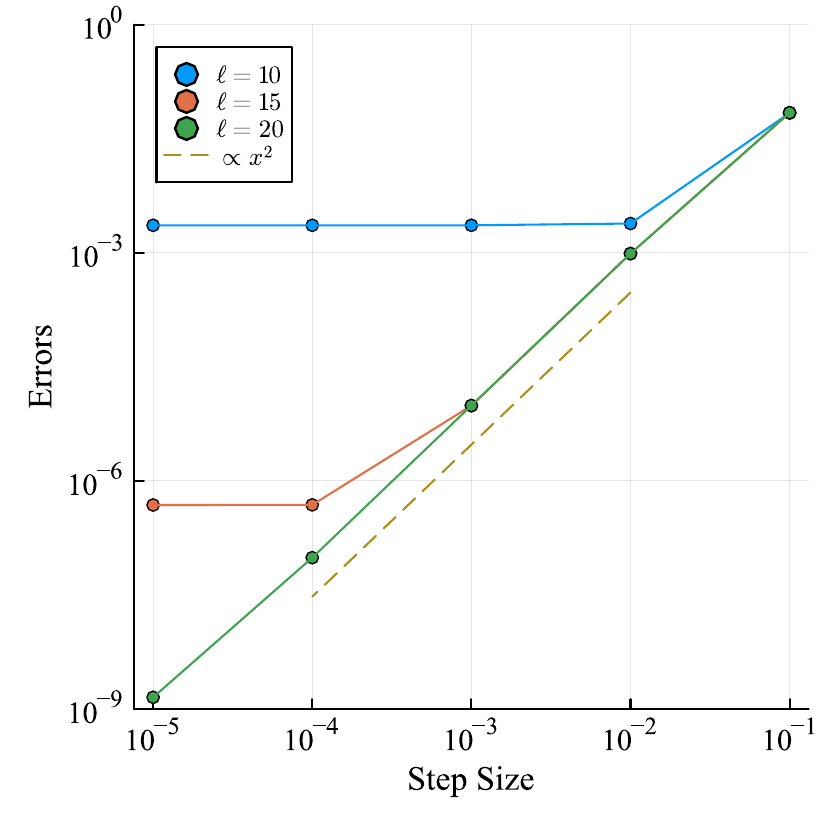}
        \caption{}
        \label{fig:errorconv_S}
    \end{subfigure}
    \caption{Figure \textbf{(a)} shows obtained evaluation time for computing the gradient of vibrational entropy in \eqref{eq:entropy} with respect to a displacement vector $u$ of length $m$. Both methods used contour integral quadrature with $\ell=20$ quadrature points. Figure \textbf{(b)} shows convergence of a second order centered finite difference computation of the gradient in \eqref{eq:entropy} for a bulk Silicon example to a gradient computed by reverse AD through the contour integral with conformal map using $\ell$ quadrature points. }
\end{figure}

\subsection{Implementation Details}
In the numerical experiments that follow we explore minimum free energy pathways for various defects in Silicon using the differentiable Stillinger-Weber interatomic potential implementation in \texttt{EmpiricalPotentials.jl} \cite{gitEP}. Forward mode AD is supported in {\tt EmpiricalPotentials.jl} via the \texttt{ForwardDiff.jl} \cite{RevelsLubinPapamarkou2016} package but manual adjoints had to be added for the Stillinger-Weber potential to allow for \texttt{Zygote.jl}~\cite{Zygote.jl-2018} and \texttt{ChainRules.jl}~\cite{FramesWhite2024} support. The search for minimum free energy pathways is performed using the \texttt{SaddleSearch.jl} \cite{gitSaddle} package which supports string and NEB based searches. The Stillinger-Weber potential in our examples is employed purely for the sake of convenience; any interatomic potential supporting the \texttt{ForwardDiff.jl} and \texttt{ChainRules.jl} interfaces can be used instead.

The source code used to generate the results is available in the companion code repository \cite{gitEntropyGrad}. Animations of the migration processes are made available via FigShare, see \cite{TorabiFigshare2024}, providing a visual representation of these dynamic effects. Throughout our experiments all atoms in the lattice are allowed to move during defect migration. 

\subsection{Complexity and Convergence Study}

To demonstrate the computational complexity and convergence behavior of evaluating the derivative of entropy $\frac{\partial \mathcal{S}(u)}{\partial u}$ using our implementation, we conducted tests on a system of silicon comprising 64 atoms. Figure \ref{fig:complexityplot_S} illustrates the CPU time complexities for reverse mode and forward mode automatic differentiation (AD) based on the proposed contour integral formulation. The observed complexity for forward and reverse mode AD, for a fixed number of quadrature points, $\ell$ scale as \(O(m^4)\) and \(O(m^2)\) respectively. Figure \ref{fig:errorconv_S} showcases the error between a simple second order centered finite difference approach and the value given by reverse AD through the contour integral approach. The FD method effectively serves as a proxy for the true gradient up to approximately $10^{-10}$, beyond which numerical precision errors limit its reliability. These observations allow one to assess the contour integral method's accuracy as a function of $\ell$ and emphasize the importance of selecting an appropriate number of quadrature points. These results are fully consistent with our numerical study on toy problems in the previous section.

\subsection{Vacancy Migration}\label{ex:vac}
\begin{figure}[!htb]
    \centering
    \begin{subfigure}[b]{0.45\textwidth} 
        \centering
        \begin{minipage}[b]{0.45 \textwidth}
    \includegraphics[width=\textwidth]{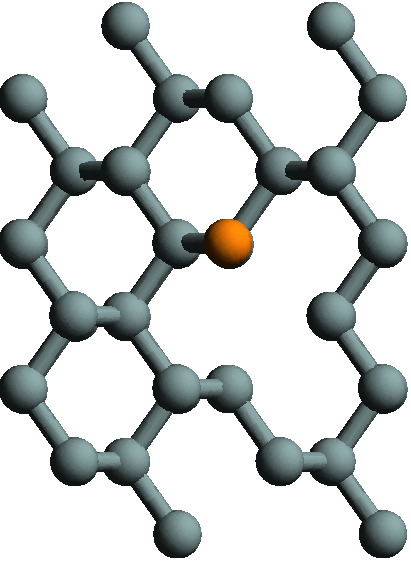}
    \caption*{(1)}
        \end{minipage}
        \hspace{0.2in}
        \begin{minipage}[b]{0.45
\textwidth}
    \includegraphics[width=\textwidth]{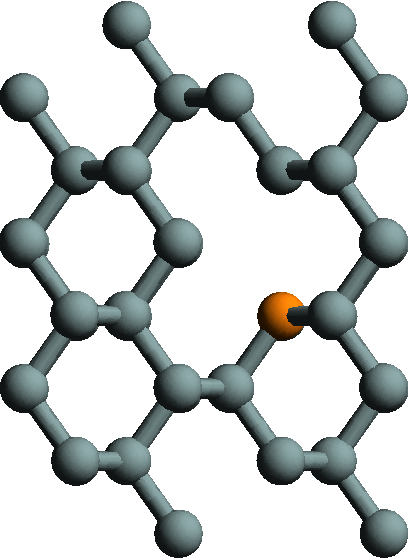}
    \caption*{(2)}
        \end{minipage}
        \caption{} 
        \label{fig:vac_a}
    \end{subfigure}
    \hspace{0.5in}
    \begin{subfigure}[b]{0.45\textwidth}
        \centering
    \includegraphics[width=\textwidth]{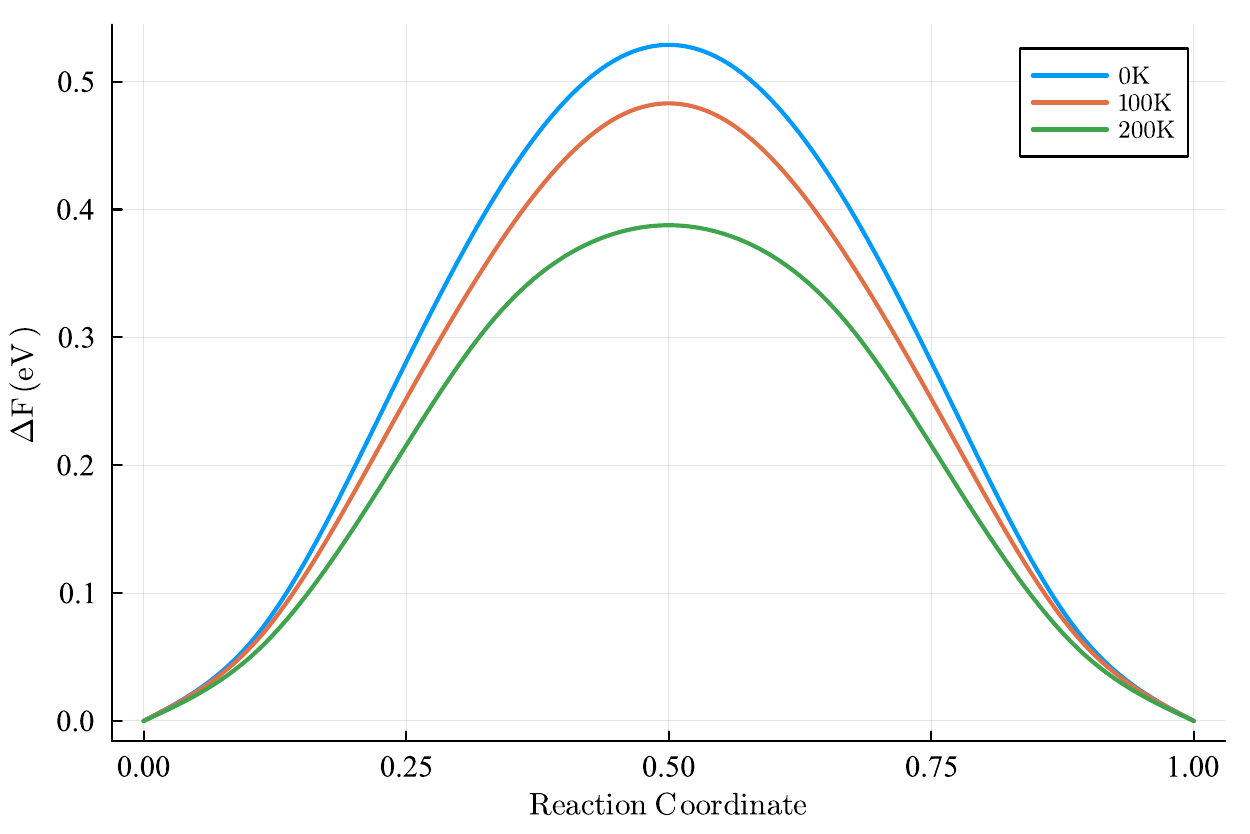}
        \caption{}
        \label{fig:vac_b}
    \end{subfigure}
    \caption{\textbf{(a)} Illustration of vacancy migration in silicon. The orange atom in (1) migrates to occupy the adjacent vacancy site, resulting in configuration (2). \textbf{(b)} Free-energy profiles for vacancy migration in silicon at various temperatures, illustrating the influence of temperature on the migration barrier and the entropic contributions that facilitate the migration process.}
    \label{fig:vacmig}
\end{figure}

Vacancies in silicon remain an extensively investigated point defect in semiconductor physics, due to their role in facilitating impurity diffusion. 
Here we consider vacancy migration, schematically illustrated in Figure \ref{fig:vacmig}, which alleviates internal stresses and influences the material's mechanical strength and ductility \cite{Coleman_2011}. 

We begin by generating a diamond cubic bulk silicon system containing 64 atoms per supercell. To create the initial configuration, one atom is removed, introducing a vacancy. The system is then relaxed by minimizing its energy to reach equilibrium.
The free-energy profiles for vacancy migration at various temperatures is illustrated in Figure  \ref{fig:vac_b}. The migration path connects the minimized initial and final configurations on the PES. Free-energy profiles were calculated at three different temperatures: 0K, 100K, and 200K. The 0K pathway corresponds to the minimum energy path (MEP) on the potential energy surface (PES). The vacancy migration energy at 0K is calculated to be 0.515 eV, which is in agreement with the literature value of 0.52 eV reported by \cite{PhysRevB.88.195204}. Our computed migration path at 0K also aligns closely with the one reported in~\cite{PhysRevB.78.035208}.

As the temperature increases, the energy barrier shows a significant variation, reflecting the dynamic nature of atomic movements and their influence on the vacancy migration pathway. 

\subsubsection{Interstitial}

\begin{figure}[!htb]
    \centering
    \begin{subfigure}[b]{0.35\textwidth}
        \centering
\includegraphics[width=\textwidth]{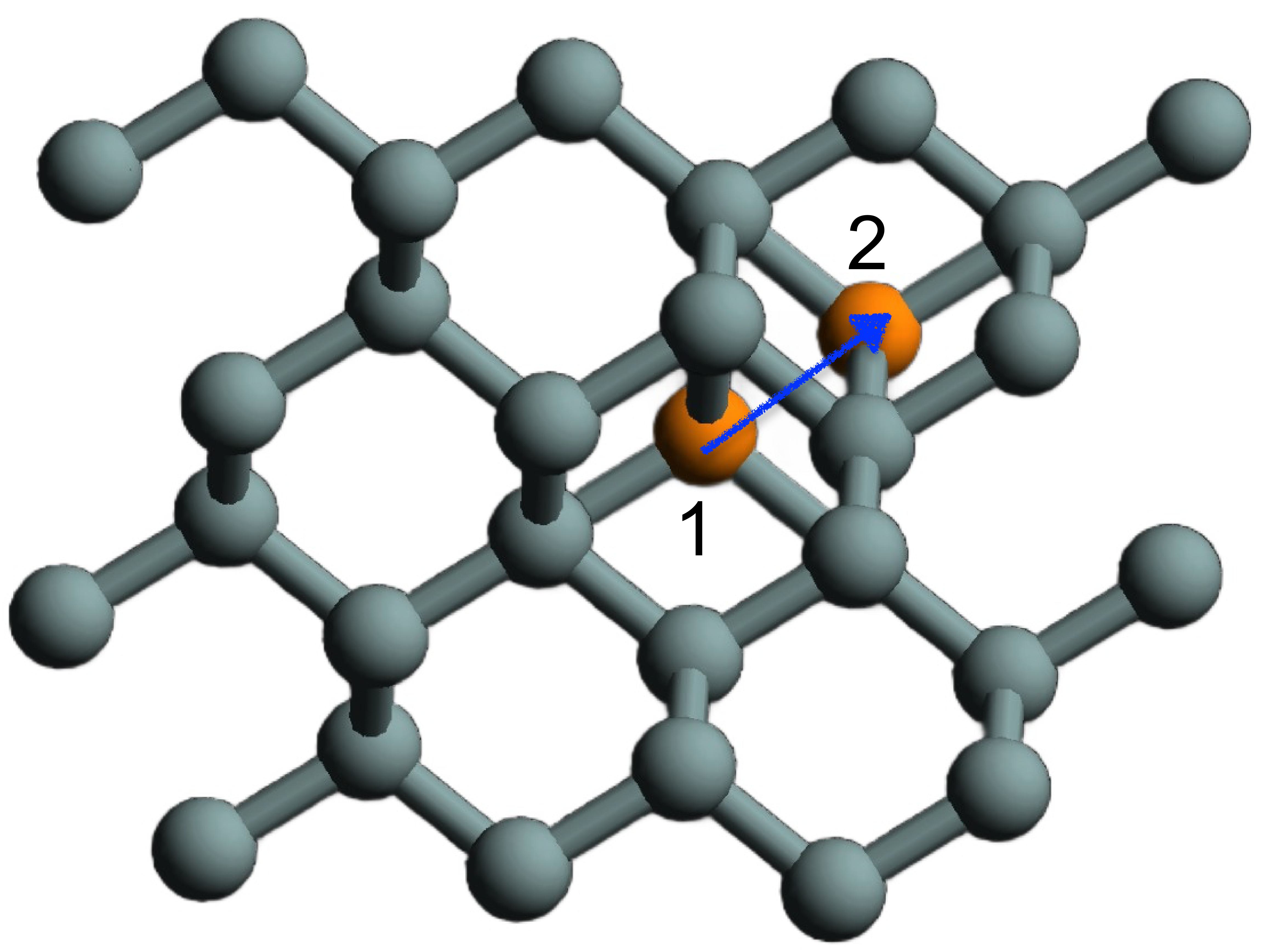}
        \caption{}
    \end{subfigure}
    \hspace{0.5in}
    \begin{subfigure}[b]{0.46\textwidth}
        \centering
\includegraphics[width=\textwidth]{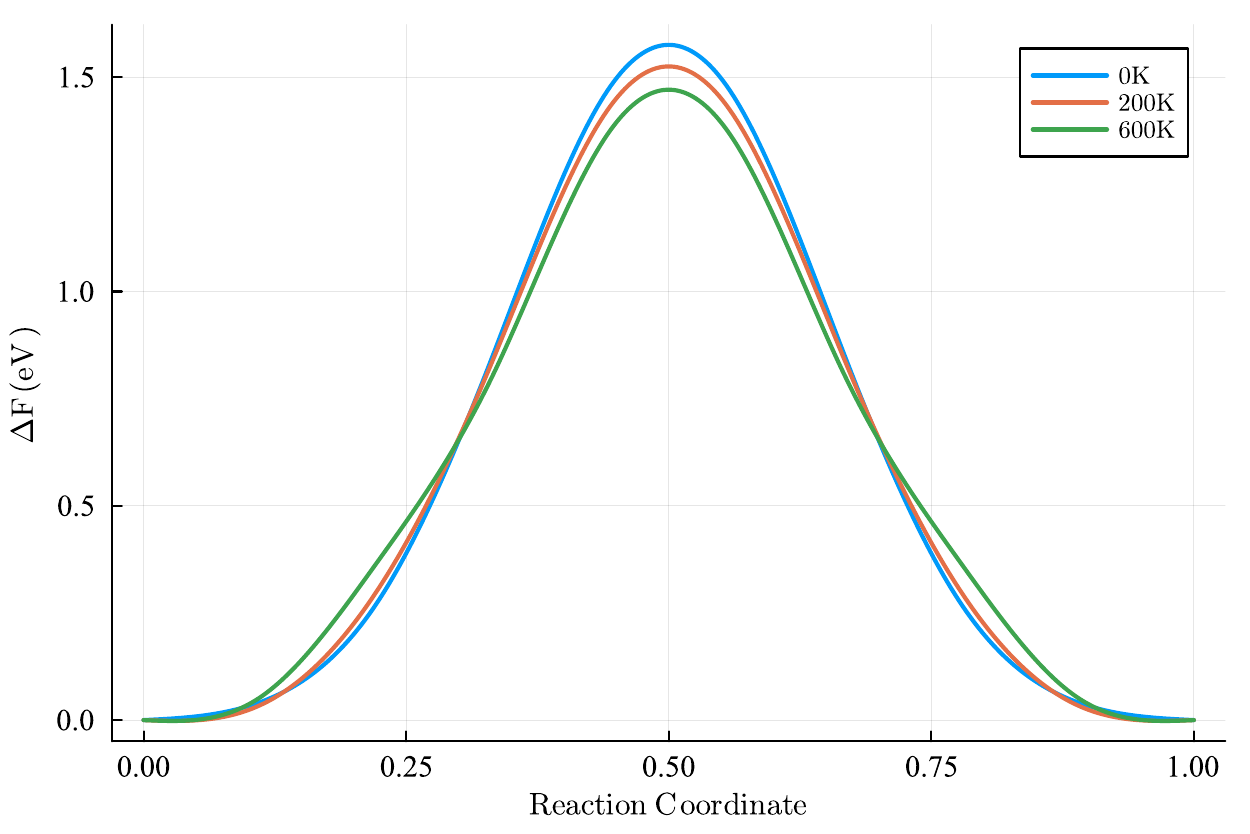}
        \caption{}
    \end{subfigure}
    \caption{\textbf{(a)} Illustration of interstitial migration in silicon, with the red atom indicating the interstitial defect migrating from position 1 to 2. \textbf{(b)} Evolution of the free-energy profiles for the migration of an interstitial defect in Silicon.}
    \label{fig:intermig}
\end{figure}

Self-interstitials in crystalline Si introduce much stronger local distortions in the crystal structure than vacancies. Interstitials typically occupy high-symmetry sites such as the tetrahedral or hexagonal interstitial sites. The migration of an interstitial atom from one site to an adjacent site, showcased in Figure \ref{fig:intermig}, is a process that influences material properties such as diffusivity and mechanical strength.

We generated a bulk silicon system consisting of 64 atoms per supercell and generated an initial state with a tetrahedral interstitial defect by inserting an atom at \(\left(\frac{a}{2}, \frac{a}{2}, \frac{a}{2}\right)\), where \(a\) is the lattice constant. The final state has an interstitial inserted at an adjacent tetrahedral site. To computed the migration path between these two states by applying the NEB method to the FES, to obtain free energy profiles at temperatures of 0K, 200K, and 600K. Figure~\ref{fig:intermig} illustrates that the free energy barrier decreases slightly with increasing temperature. Thermal effects reduce the energy barrier, making interstitial migration slightly more likely at higher temperatures.

\subsubsection{Variational TST}\label{sec:vtst}
Understanding the rates of various thermally activated processes, ranging from defect formation \cite{bagchi2024anomalousentropydrivenkineticsdislocation} and migration \cite{865257} to creep \cite{10.1007/978-3-319-48254-5_42} and catalysis \cite{doi:10.1021/jacs.3c04030}, is crucial for the advancement and optimization of material properties and chemical processes. These rates are often calculated using harmonic transition state theory (HTST), where the potential energy of the lattice is approximated by a second-order Taylor expansion, effectively assuming a quadratic potential near equilibrium positions. HTST can yield inaccurate predictions when the potential energy surface significantly deviates from this quadratic assumption or when thermal effects substantially influence the system's dynamics \cite{PhysRevLett.120.135503}. As such, a more precise approach involves including non-harmonic corrections to resolve additional complexity found in the system's potential landscape.

Once the minimum energy path (MEP) for a reaction is determined, the transition rate can be defined within the framework of harmonic transition state theory (HTST) as:

\begin{equation}
\mathcal{K}^{\rm HTST} = \exp\left(-\beta \left[\mathcal{F}(\xi_{\text{saddle}}) - \mathcal{F}(\xi_{\text{min}})\right]\right), \label{eq:HTST}
\end{equation}
where \( \xi_{\text{min}} \) denotes the positions corresponding to the minimum energy configuration, and \( \xi_{\text{saddle}} \) represents the positions of the saddle configuration. The MEP can be parameterized by a reaction coordinate \( \xi \), a scalar variable that describes the progress of the system along the reaction path. If \( u(\xi) \) represents the reaction path, then \( \xi \) serves as a continuous parameter ranging from initial minimum energy configuration (\( \xi = 0 \)) to the final minimum configuration (\( \xi = 1 \)). The reaction coordinate provides a natural way to track changes in energy and other system properties along the MEP. The saddle point is identified as the maximum energy point along the reaction coordinate. Here, \( \mathcal{F} \) denotes the free energy, and \( \beta = 1 / (k_B T) \), where \( k_B \) is the Boltzmann constant and \( T \) is the temperature.

To achieve a more accurate description of defect migration and formation in silicon, one can incorporate higher-order terms in the transition state formulation. These additional terms introduce corrections to both the energy and entropy contributions, impacting the calculated transition rates and leading to a more precise and reliable model of defect dynamics. Variational transition state theory (VTST) \cite{bagchi2024anomalousentropydrivenkineticsdislocation, C7CS00602K} provides a simple correction of this kind. Below, we briefly review the idea.

One can equivalently express the energy \( \mathcal{E} \) and entropy \( \mathcal{S} \) as a function of the reaction coordinate \( \xi \) and expand both along the reaction coordinate on the MEP around the saddle point \( \xi = \xi_{\text{saddle}} \):
\begin{align}
\mathcal{E}(\xi_{\text{saddle}} + \xi) &\approx \mathcal{E}(\xi_{\text{saddle}}) + \frac{1}{2} \mathcal{E}_{\xi\xi}(\xi_{\text{saddle}}) \xi^2, \\
\mathcal{S}(\xi_{\text{saddle}} + \xi) &\approx \mathcal{S}(\xi_{\text{saddle}}) + \mathcal{S}_\xi(\xi_{\text{saddle}}) \xi,
\end{align}

where $\mathcal{E}_{\xi\xi} = \displaystyle\frac{\partial^2 \mathcal{E}}{\partial \xi^2}$ and $\mathcal{S}_{\xi} = \displaystyle\frac{\partial\mathcal{S}}{\partial \xi}$ are scalar functions. We can formally approximate 
\begin{equation}
\mathcal{F}(\xi) \approx \mathcal{E}(\xi_{\text{saddle}}) + \frac{1}{2} \mathcal{E}_{\xi\xi}(\xi_{\text{saddle}}) \xi^2 - T \mathcal{S}(\xi_{\text{saddle}}) - T \mathcal{S}_\xi(\xi_{\text{saddle}}) \xi.
\end{equation}


Since we have a maximum at the saddle point along the reaction coordinates 
on the MEP, we take the derivative of the free energy with respect to \(\xi \), the displacement along the reaction coordinates. This leads to the following condition for the stationary point:


\begin{equation}
\frac{\partial \mathcal{F}}{\partial \xi} = \mathcal{E}_{\xi\xi}(\xi_{\text{saddle}}) \xi - T \mathcal{S}_\xi (\xi_{\text{saddle}}) = 0,
\end{equation}

with solution 
\begin{equation}
\xi = T \frac{\mathcal{S}_{\xi}(\xi_{\text{saddle}})}{\mathcal{E}_{\xi\xi}(\xi_{\text{saddle}})}.
\end{equation}

Substituting this back into the free energy expression, we get:
\begin{align}
\mathcal{F}(T) \approx \mathcal{E}(\xi_{\text{saddle}}) - T \mathcal{S}(\xi_{\text{saddle}}) - \frac{1}{2} T^2 \frac{\mathcal{S}^2_{\xi}(\xi_{\text{saddle}})}{\mathcal{E}_{\xi\xi}(\xi_{\text{saddle}})}.
\end{align}
Thus,
\begin{equation}
\mathcal{F}_{vTST}(T) \approx \mathcal{F}_{HTST}(T) - \frac{1}{2} T^2 \frac{\mathcal{S}^2_{\xi}(\xi_{\text{saddle}})}{\mathcal{E}_{\xi\xi}(\xi_{\text{saddle}})}.
\end{equation}

Therefore, the temperature-dependent variational transition rate is:

\begin{equation}
\begin{aligned}
    k_{vTST}(T) &= \exp(-\beta \mathcal{F}_{HTST}) \exp\left(\beta T^2 \frac{\mathcal{S}^2_{\xi}(\xi_{\text{saddle}})}{2\mathcal{E}_{\xi\xi}(\xi_{\text{saddle}})}\right) \\
&= k_{HTST} \exp\big(\beta T^2 T_{e}) 
\end{aligned}
\end{equation}
where $T_e = \frac{\mathcal{S}^2_{\xi}(\xi_{\text{saddle}})}{2\mathcal{E}_{\xi\xi}(\xi_{\text{saddle}})}$, and is referred to as the "effective temperature".
As shown in previous sections, $\mathcal{S}_{\xi}$  can be easily obtained using our differentiation framework
without the need to rely on approximation or sampling techniques which are often used for this evaluation~\cite{bagchi2024anomalousentropydrivenkineticsdislocation}. To compare the formation rates estimated using harmonic and variational TST, we examine the migration of vacancy and interstitial defects based on the free energy profiles discussed in the previous subsections.

\begin{figure}
     \centering
        \includegraphics[width=0.5\textwidth]{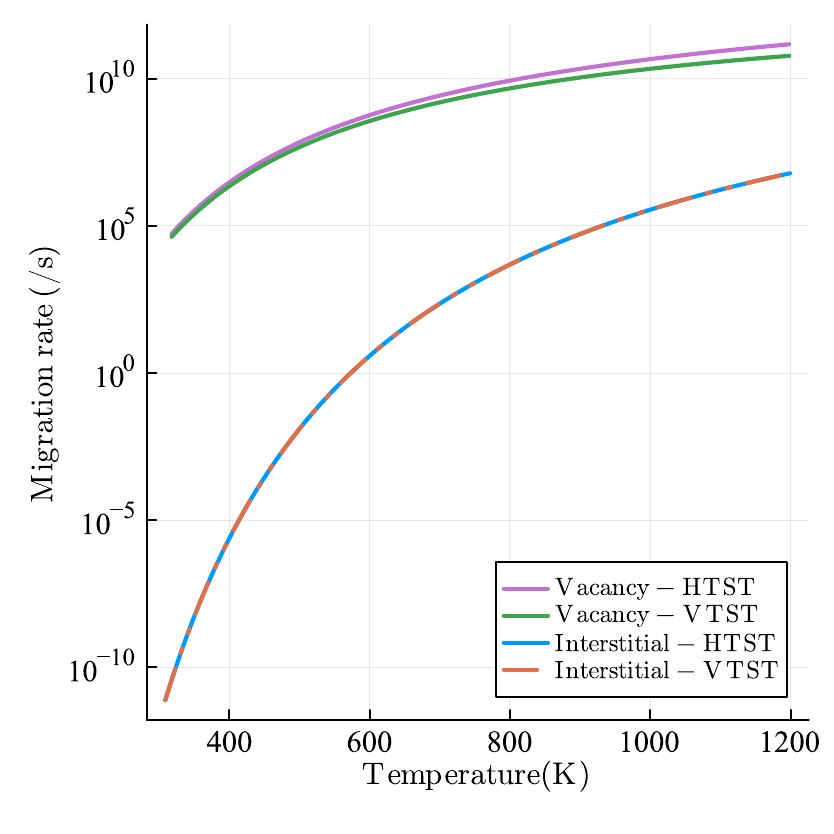}
    \caption{Estimated migration rates obtained from the variational
free energy barriers for vacancy and interstitial migration.} 
    \label{fig:vtst}
\end{figure}

Figure~\ref{fig:vtst} presents the migration rates obtained using harmonic transition state theory (HTST) and variational transition state theory (VTST). For vacancy migration, as the temperature increases, the difference between the predictions of VTST and HTST becomes more apparent. This indicates that harmonic TST overestimates the migration rate and is, therefore, less reliable for high-temperature processes.
This is evident from the fact that \( \mathcal{E}_{ww}(\xi_{\text{saddle}}) < 0 \), leading to 
\[
\exp\bigg(-\beta \frac{T^2}{2} T_e \bigg) < 1.
\]
In contrast, for interstitial migration, the temperature correction \( T_e \) is small, and the overestimation of the rate by HTST is negligible even at elevated temperatures.

For certain systems entropic contributions become increasingly significant at elevated temperatures as the vibrational modes of atoms exhibit greater anharmonicity. This leads to a wider distribution of energy states that harmonic TST can fail to capture. More advanced methods such as VTST can be used in such circumstances to obtain more reliable predictions of migration rates.

\section{Conclusion}
In this paper we showed how to accurately and efficiently compute the derivatives of functions of matrix families $X_\theta$ and their derivatives with respect to the entries of $\theta_j$ of $\theta$. The method is an extension of an established approach for matrix functions due to Hale, Higham and Trefethen \cite{doi:10.1137/070700607} and thus begins with the Cauchy integral definition of matrix functions and computes appropriate conformal maps and quadrature in the complex plane. We explained in detail the resulting computational complexity for both reverse and forward mode differentiation.

With the extension to derivatives in hand, we demonstrated the utility of the approach in natural applications in the context of molecular simulations and material modelling. We considered questions involving the entropy of Silicon systems which require repeated computation of matrix trace-logarithms and gradients thereof for saddle point searches. As the conformal map based method allows defining straightforward adjoints for the matrix logarithm, all that is required in principle for efficient reverse mode automatic differentiation of such matrix functions is a reverse mode compatible implementation of the involved interatomic potential (in the case of our Silicon example this role is played by the Stillinger-Weber potential).

Our aim was to demonstrate that employing modern differentiation algorithms not only results in numerically robust and accurate derivatives but also in improved computational performance, and that such an approach is entirely practical within a modern software stack.

\section{Acknowledgments}

TT and CO were supported by NSERC Discovery grant RGPIN-2021-03489 and NFRF Exploration Grant GR022937. TSG was partially supported by a PIMS-Simons postdoctoral fellowship, jointly funded by the Pacific Institute for the Mathematical Sciences (PIMS) and the Simons Foundation.

\newpage
\section{Appendix}
\subsection{Jacobi Elliptic Functions} \label{app:JEF}
We begin with some definitions and properties of Jacobi elliptic functions as e.g. detailed in \cite{AbramowitzStegun1965, ByrdFriedman1971}.
\subsubsection{Elliptic Functions}
A \emph{doubly periodic meromorphic function} is called an \emph{elliptic function}. Let a \emph{parameter} \( m \) and \emph{complementary parameter} \( p \) be given satisfying
\begin{equation}
m + p = 1.
\end{equation}
If \( m \in \mathbb{R} \) then in what follows one can assume without loss of generality that \( 0 \leq m \leq 1 \).
\subsubsection{Quarter-Periods}
The \emph{quarter-periods} \( K \) and \( iK' \) are defined by the integrals
\begin{align}
K(m) &= K = \int_0^{\frac{\pi}{2}} \frac{d\theta}{\sqrt{1 - m \sin^2 \theta}}, \\
iK'(m) &= iK' = \int_0^{\frac{\pi}{2}} \frac{d\theta}{\sqrt{1 - p \sin^2 \theta}},
\end{align}
where \( K, K' \in \mathbb{R} \). \( K \) is called the real quarter-period and \( iK' \) the imaginary quarter-period. These quantities satisfy the relationships
\begin{equation}
K(m) = K(p) = K'(1 - m).
\end{equation}
Moreover, if any one of the numbers \( m, p, K(m), K'(m), \frac{K'(m)}{K(m)} \) are given, all the rest are uniquely determined through the above relations, i.e. \( K \) and \( K' \) cannot be independently chosen.
\subsubsection{Jacobi Elliptic Functions}
The Jacobi elliptic functions are widely studied standard forms of elliptic functions represented by \( \text{sn}(u, k) \), \( \text{cn}(u, k) \), and \( \text{dn}(u, k) \), where \( k \) is termed the elliptic modulus. They originate from the inverse of the elliptic integral of the first kind,
\begin{equation}
u = F(\varphi, k) = \int_0^{\varphi} \frac{dt}{\sqrt{1 - k^2 \sin^2 t}},
\end{equation}
where \( \varphi \) as \( \text{am}(u, k) \) denotes the Jacobian amplitude. This leads to the relationships
\begin{align}
\sin \varphi &= \sin(\text{am}(u, k)) = \text{sn}(u, k), \\
\cos \varphi &= \cos(\text{am}(u, k)) = \text{cn}(u, k), \\
\sqrt{1 - k^2 \sin^2 \varphi} &= \sqrt{1 - k^2 \sin^2(\text{am}(u, k))} = \text{dn}(u, k).
\end{align}
These functions extend trigonometric functions to be doubly periodic, satisfying:
\begin{align}
\text{sn}(u, 0) &= \sin u, \\
\text{cn}(u, 0) &= \cos u, \\
\text{dn}(u, 0) &= 1.
\end{align}

\subsubsection{Identities for Jacobi Elliptic Functions}
The established identities for Jacobi elliptic functions are given by the following equations:
\begin{align}
\text{sn}^2 u + \text{cn}^2 u &= 1, \\
k^2 \text{sn}^2 u + \text{dn}^2 u &= 1, \\
k^2 \text{cn}^2 u + k'^2 &= \text{dn}^2 u, \\
\text{cn}^2 u + k'^2 \text{sn}^2 u &= \text{dn}^2 u.
\end{align}
\subsubsection{Specific Values}
Specific noteworthy values are listed below:
\begin{align}
\text{cn}(0, k) &= \text{cn}(0) = 1, \\
\text{cn}(K(k), k) &= \text{cn}(K(k)) = 0, \\
\text{dn}(0, k) &= \text{dn}(0) = 1, \\
\text{dn}(K(k), k) &= \text{dn}(K(k)) = k' = \sqrt{1 - k^2}, \\
\text{sn}(0, k) &= \text{sn}(0) = 0, \\
\text{sn}(K(k), k) &= \text{sn}(K(k)) = 1,
\end{align}
where \( K = K(k) \) signifies the complete elliptic integral of the first kind and \( k' = \sqrt{1 - k^2} \) represents the complementary elliptic modulus \cite{Whittaker_Watson_1996}.

\subsubsection{Complex Arguments}
When dealing with complex arguments, the Jacobi elliptic functions can be extended as follows:
\begin{align}
\text{sn}(u + iv) &= \frac{\text{sn}(u, k) \text{dn}(v, k')}{1 - \text{dn}^2(u, k) \text{sn}^2(v, k')} + \frac{i \text{cn}(u, k) \text{dn}(u, k) \text{sn}(v, k') \text{cn}(v, k')}{1 - \text{dn}^2(u, k) \text{sn}^2(v, k')}, \\
\text{cn}(u + iv) &= \frac{\text{cn}(u, k) \text{cn}(v, k')}{1 - \text{dn}^2(u, k) \text{sn}^2(v, k')} - \frac{i \text{sn}(u, k) \text{dn}(u, k) \text{sn}(v, k') \text{dn}(v, k')}{1 - \text{dn}^2(u, k) \text{sn}^2(v, k')}, \\
\text{dn}(u + iv) &= \frac{\text{dn}(u, k) \text{cn}(v, k') \text{dn}(v, k')}{1 - \text{dn}^2(u, k) \text{sn}^2(v, k')} - \frac{i k^2 \text{sn}(u, k) \text{cn}(u, k) \text{sn}(v, k')}{1 - \text{dn}^2(u, k) \text{sn}^2(v, k')}.
\end{align}
\begin{figure}[!htb]
    \centering
    \includegraphics[width=0.4\textwidth]{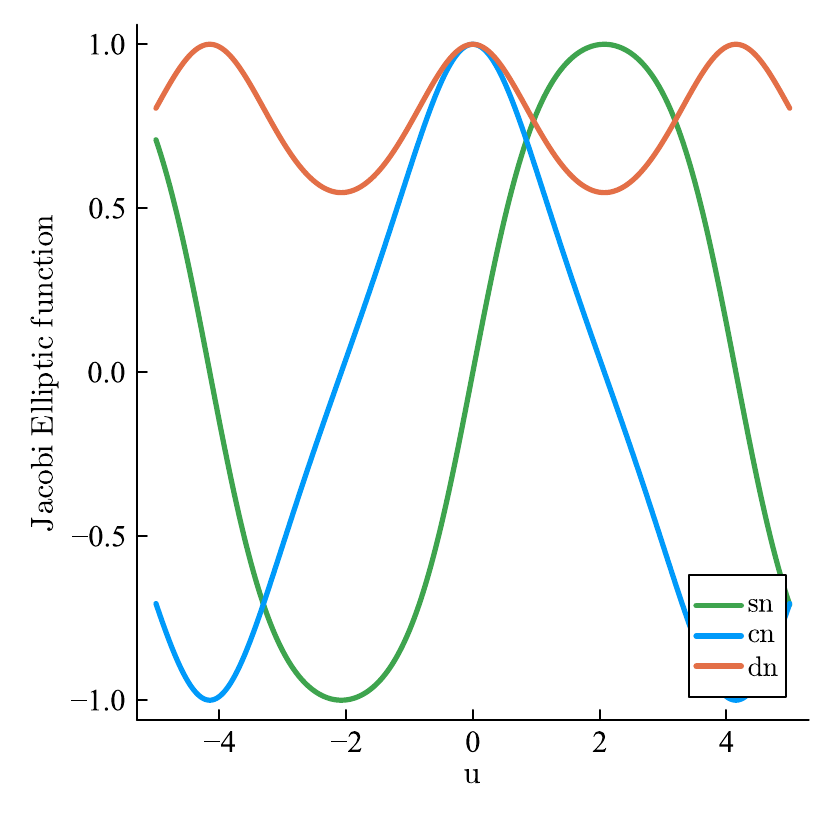}
    \caption{Jacobi Elliptic Functions \( \text{sn}(u, k) \), \( \text{cn}(u, k) \), \( \text{dn}(u, k) \) for \( m = 0.7 \).}
    \label{fig:jef}
\end{figure}
\newpage
\subsection{Strategies for Utilizing Matrix Sparsity} \label{app:coloring}
Leveraging potential sparsity of the matrix \( X \) can significantly enhance the efficiency of computing its Jacobian. Consider the toy model presented in Section~\ref{sec:jacobian}: 
\begin{equation}
E(C; u) = \sum_{i} \left( \sum_{j \in \mathcal{N}_i} C_{ij} |u_i - u_j|^2 + \delta |u_i - u_j|^3 + |u_i - u_j|^4 \right).
\end{equation}

\begin{figure}[!htb]
     \centering
        \includegraphics[width=0.8\textwidth]{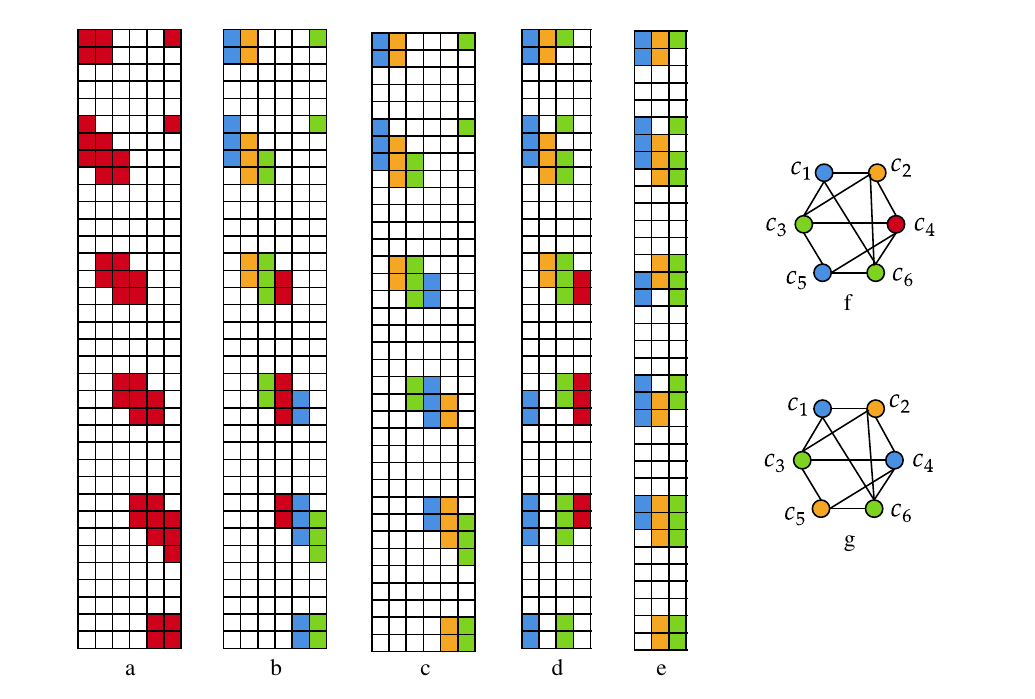}
    \caption{The process of compressing the Jacobian matrix of the Hessian of toy model~\ref{eq:toymodel} with 8 atoms in two partitions (a)–(e). Each partition is also represented as a distance-1 coloring in a column intersection graph (f)-(g).} 
    \label{fig:Jac_color}
\end{figure}

Consider a system composed of 8 atoms. Our objective is to compute the Jacobian of the Hessian matrix, which corresponds to the matrix of second derivatives of the energy \ref{eq:toymodel} with respect to the displacements in such a system.
 The Jacobian is shown in Figure \ref{fig:Jac_color} (a). Looking at this sparse matrix, we can see that there is a lot of free space in the matrix. We aim to compress the sparse matrix into a denser format shown in Figures \ref{fig:Jac_color} (b) and (c), allowing for the processing of non-zero entries with significantly fewer function calls than previously necessary. Although both matrices (d) and (e) represent the Jacobian in a denser format, the main task is to find the smallest dense matrix which in our case is (e). This technique uses a strategy from graph theory \cite{SparseJac, kubale2004graph}, aiming to combine columns with non-overlapping non-zero elements, referred to as {\it structurally orthogonal columns} into single groups, thus reducing the total number of groups needed. To achieve this, one can use a number of graph coloring algorithms including Contraction Coloring, Greedy distance-k coloring, and Backtracking Sequential Coloring. We will briefly go over the Greedy distance-1 coloring algorithm which results were also shown in figure \ref{fig:Jac_color} (f)-(g)~\cite{Torabi_2024}.

We begin with a brief overview of essential graph theory terminology. A graph \( G \) is formally defined as an ordered pair \( G = (V, E) \), where \( V \) is a finite, non-empty set of vertices, and \( E \) is a set comprising unordered pairs of distinct vertices, known as edges. Vertices \( u \) and \( v \) are adjacent if an edge \( (u, v) \) is included in \( E \); otherwise, they are described as non-adjacent. Within a graph, a path of length \( l \) (measured in edges) is a sequence of vertices \( v_1, v_2, \ldots, v_{l+1} \), where each consecutive pair \( (v_i, v_{i+1}) \) is adjacent, for all \( 1 \leq i \leq l \), with each vertex appearing uniquely in the sequence. Vertices \( u \) and \( v \) are considered distance-\( k \) neighbors if the shortest path between them has a length of \( k \) or less. The set of all distance-\( k \) neighbors of a vertex \( u \), denoted \( N_k(u) \), does not include \( u \) itself. Additionally, if two vertices are distance-\( k \) neighbors, they also qualify as distance-\( k' \) neighbors for any \( k' > k \).

A graph is categorized as bipartite if its vertex set \( V \) can be partitioned into two disjoint subsets \( V_1 \) and \( V_2 \), such that every edge connects a vertex from \( V_1 \) to one from \( V_2 \). There are no edges between vertices within the same subset, which ensures that \( V_1 \) and \( V_2 \) comprehensively separate the vertices of the graph.

A distance-\( k \) vertex coloring of a graph \( G = (V, E) \) is a labeling function \( \varphi: V \to \{1, 2, \ldots, p\} \) that assigns different colors to any pair of distance-\( k \) neighbors. The minimum number of colors needed to establish such a coloring for graph \( G \) is termed the \( k \)-chromatic number, denoted as \( \chi_k(G) \). If only a specific subset \( W \subset V \) of the vertices is colored, the coloring is referred to as partial. Specifically, a partial distance-\( k \) coloring on \( W \) is defined by a function \( \varphi: W \to \{1, 2, \ldots, p\} \) such that \( \varphi(u) \neq \varphi(v) \) for any two vertices \( u \) and \( v \) within \( W \) that are distance-\( k \) neighbors. An example of such a distance-2 coloring is illustrated on the left side of Figure 2.1.

The following lemma provides a graph-theoretical characterization of structural orthogonality for a nonsymmetric matrix.

\begin{lemma}{\cite[Lemma 3.4]{SparseJac}} \label{lem:k-dis}
Let $A$ be an $m \times n$ matrix and $G_b(A) = (V_1, V_2, E)$ be its bipartite graph. Two columns (or rows) in $A$ are structurally orthogonal if and only if the corresponding vertices in $G_b(A)$ are at a distance greater than two from each other.
\end{lemma}
According to Lemma \ref{lem:k-dis}, determining a structurally orthogonal partition of the columns of a matrix \(A\) corresponds to attaining a partial distance-2 coloring of its bipartite graph \(G_b(A) = (V_1, V_2, E)\) specifically applied to \(V_2\). This coloring is termed partial as it does not extend to \(V_1\). Theorem 3.5 provides a formal statement of this relationship.

\begin{theorem}{\cite[Theorem 3.5]{SparseJac}} Let \(A\) be an \(m \times n\) matrix and \(G_b(A) = (V_1, V_2, E)\) represent its bipartite graph. A function \(\varphi\) constitutes a partial distance-2 coloring of \(G_b(A)\) on \(V_2\) if and only if \(\varphi\) generates a structurally orthogonal partition of the columns of \(A\).
\end{theorem}
Now we can reformulate our main problem as follows:
\textit{Given the bipartite graph \(G_b(A) = (V_1, V_2, E)\) that represents the sparsity structure of an \(m \times n\) matrix \(A\), the objective is to find a partial distance-2 coloring of \(G_b(A)\) on \(V_2\) that utilizes the minimal number of colors. This coloring strategy aims to efficiently partition the columns of \(A\) into structurally orthogonal sets, corresponding to the vertices in \(V_2\), while minimizing the color count.}

To solve this problem we can use the greedy distance-2 coloring algorithm, shown in \ref{alg:greed2}. We initially need to represent our Jacobian using a graph. Given a matrix \(A \in \mathbb{R}^{m \times n}\), we can represent the structure of \(A\) using a bipartite graph \(G = (V, E)\). The construction process for the bipartite graph is outlined as follows:

\begin{enumerate}
    \item Create two disjoint sets of vertices, \(U\) and \(W\), where \(U\) corresponds to the rows of \(A\) and \(W\) corresponds to the columns of \(A\). Thus, \(U = \{u_1, u_2, \ldots, u_m\}\) and \(W = \{w_1, w_2, \ldots, w_n\}\).
    \item The set of vertices \(V\) is the union of \(U\) and \(W\), i.e., \(V = U \cup W\).
    \item For each non-zero entry \(A_{ij}\) in the matrix \(A\), add an edge \((u_i, w_j)\) to the set of edges \(E\). This implies there is an edge between vertex \(u_i \in U\) and vertex \(w_j \in W\) if and only if the entry \(A_{ij}\) is non-zero.
\end{enumerate}

This bipartite graph representation, \(G\), captures the interactions between the rows and columns of \(A\) based on its non-zero entries. It allows for a visual and analytical understanding of the matrix's structure, and can help us use the coloring algorithms that exploit sparsity patterns for improved computational efficiency. Below, we outline the detailed steps and the corresponding algorithm that collectively streamline this process:

\begin{enumerate}
    \item \textbf{Graph Representation of Sparsity:}
    First, the sparsity pattern of the Jacobian matrix \(J\) is modeled as a graph \(G = (V, E)\), where the vertices \(V\) represent the matrix's rows and columns, and the edges \(E\) correspond to non-zero entries in \(J\).

    \item \textbf{Applying a Greedy Distance-2 Coloring Algorithm:}
    The coloring of graph \(G\) is executed through a methodical greedy algorithm that ensures no two vertices within two edges of each other share the same color, reflecting the distance-2 coloring strategy necessary for avoiding computational interference:
    \begin{enumerate}
        \item Initially, all vertices are uncolored.
        \item The coloring process is iterative, with each uncolored vertex being assigned the least positive integer color that is not used by its distance-1 and distance-2 neighbors, ensuring that the coloring satisfies the distance-2 constraints.
    \end{enumerate}
    
    \item \textbf{Constructing Perturbation Vectors:}
    For every unique color assigned, a corresponding perturbation vector \(d\) is created. This vector has elements set to 1 for indices colored with the current color and 0 elsewhere.

    \item \textbf{Evaluating the Jacobian:}
    Using each perturbation vector \(d\), the function \(F(x + \epsilon d)\) is evaluated to ascertain the non-zero components of the Jacobian matrix that correspond to each color. This step is crucial for identifying which parts of the matrix can be independently calculated, thereby enhancing computational efficiency.
\end{enumerate}

\begin{algorithm}
\caption{A greedy distance-2 coloring algorithm. \cite[Algorithm 3.1]{SparseJac}}
\label{alg:greed2} 
\begin{algorithmic}[1] 
\Procedure{D2ColoringAlg}{$G = (V, E)$}
    \State Let $v_1, v_2, \ldots, v_{|V|}$ be a given ordering of $V$
    \State Initialize \textbf{ForbiddenColors} with some value $a \notin V$
    \For{$i \gets 1$ to $|V|$}
        \For{each colored vertex $w \in N_1(v_i)$}
            \State $\textbf{ForbiddenColors}[color[w]] \gets v_i$
            \For{each colored vertex $x \in N_1(w)$}
                \State $\textbf{ForbiddenColors}[color[x]] \gets v_i$
            \EndFor
        \EndFor
        \State $color[v_i] \gets \min\{c > 0 : \textbf{ForbiddenColors}[c] \neq v_i\}$
    \EndFor
\EndProcedure
\end{algorithmic}
\end{algorithm} 

Algorithm \ref{alg:greed2}, operationalizes the second step of this process. Once a group is determined, centered finite difference or automatic differentiation \cite{Torabi2024} can be used to calculate the directional derivatives along the compressed matrix directions. For more details on the graph coloring method for computing derivatives we refer to~\cite{SparseJac}.

\newpage
\printbibliography[heading=bibintoc]

\end{document}